\documentclass[aps,prb,superscriptaddress]{revtex4}
\usepackage{graphicx}
\usepackage{amsmath}
\usepackage{epstopdf}
\usepackage{rotating,dcolumn}
\begin{document}
\title{Magnetic chirality as probed by neutron scattering}
\author{V. Simonet}
\email[]{virginie.simonet@grenoble.cnrs.fr}
\affiliation{Institut N\'eel, CNRS \& Universit\'e Joseph Fourier,
BP166, 38042 Grenoble Cedex 9, France}
\author{M. Loire}
\affiliation{Institut N\'eel, CNRS \& Universit\'e Joseph Fourier,
BP166, 38042 Grenoble Cedex 9, France}
\author{R. Ballou}
\affiliation{Institut N\'eel, CNRS \& Universit\'e Joseph Fourier,
BP166, 38042 Grenoble Cedex 9, France}
\begin{abstract}
We review the concept of chirality, at first briefly in a general context then in the specific framework of the spin networks. We next discuss to what extent neutron scattering appears as an unconvertible tool to probe magnetic chirality in the static and dynamical regimes of the spins. The remarkable chiral ground state and excitations of the Fe$-$langasite compound finally serves to illustrate the use of neutron polarimetry in the experimental studies of the magnetic chirality.
\end{abstract}
\maketitle

\section{Facets of Chirality}
\label{intro}

\begin{figure}
\resizebox{0.9\columnwidth}{!}
{\includegraphics{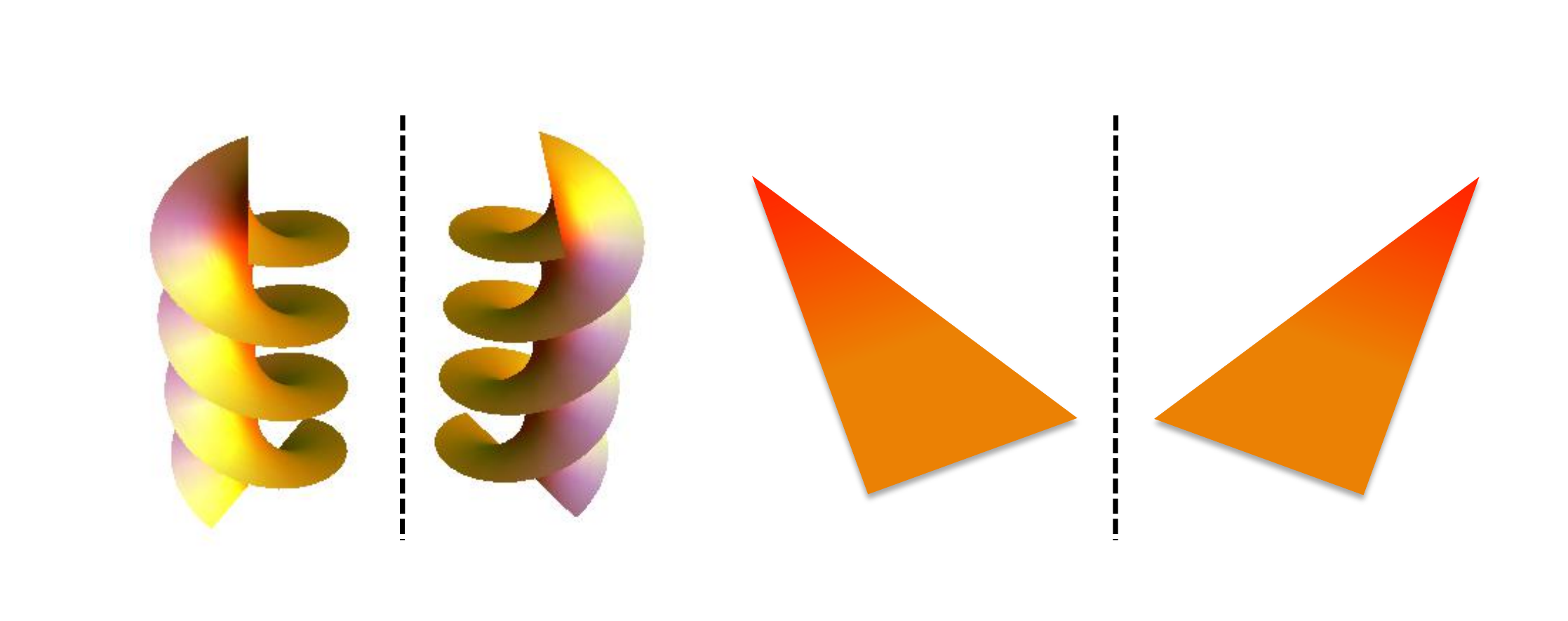} }
\caption{Examples of chirality in 3 dimensions with the two enantiomorphs of an helix (left) and in 2 dimensions with the two enantiomorphs of a non-isoceles triangle (right).}
\label{fig:1}       
\end{figure}

The word chiral, which comes from the greek $\chi\epsilon\iota\rho$ for hand, was introduced by Lord Kelvin in 1904 to describe an object whose image in a plane mirror cannot be brought in coincidence with itself \cite{Kelvin1904}. The two images are called enantiomorphs (enantiomers for molecules). An archetype is the cylindrical helix (see figure \ref{fig:1}). Pasteur was before aware of the concept, which he called "dissym\'etrie". He identified it through crystal morphology and optical activity \cite{Pasteur1884}. One of his major fulfillments had been to separate a racemic mixture (equal proportions of enantiomorphic species) of sodium ammonium paratartrate in two enantiopure subsets which were rotating the plane of polarization of a linearly polarized ingoing light by angles of equal magnitude but of opposite sign.

Chirality is a key property in {\bf chemistry and biology} \cite{Wagniere}. It is crucial to life, which basically is homochiral. A number of biological functions are activated or inhibited according to highly precise molecular mechanisms, in which chiral host molecules recognize enantiomeric guest molecules in different ways. Enantiomers often smell and taste differently. The structural difference between enantiomers can be serious with respect to the actions of synthetic drugs, resulting for instance in marked differences in the pharmacological efficiencies of enantiomers. The importance of these issues has led to the attribution of the Nobel Prize of Chemistry in 2001 to William S. Knowles, Ryoji Noyori and K. Barry Sharpless for their work on chiral catalysis. Chirality is often an emergent phenomenon, that is to say  inherent to a specific organization of sub-objects not necessarily chiral, but is also encountered at the most fundamental level. It was for instance found out that the neutrino displays only a left helicity, that is to say its spin is observed systematically antiparallel to its linear momentum. This is explained by postulating that the {\bf weak interaction} is intrinsically not invariant under mirror symmetry \cite{WeakInt}, acting only on left-chiral fermions or right-chiral anti-fermions \footnote{Chiral states of a Dirac fermion refer to the eigenstates of the operator $i ~\gamma^0 \gamma^1 \gamma^2 \gamma^3$, which has eigenvalues $\pm 1$, where $\gamma^\mu$ are the Dirac's operators defined from the anticommutation relation $\{\gamma^\mu, \gamma^\nu\} = 2 g^{\mu\nu} \mathcal{I}$ with $\mathcal{I}$ the identity operator and $g^{\mu\nu}$ the Minkowski metric. Note also that the helicity of a fermion is the same as its chirality solely in the zero mass limit. If the fermion is massive then its helicity is a frame dependent property, unlike its chirality.}. The discovery of the breaking of the mirror symmetry by the weak interactions was so unexpected that it had led to the attribution of the Nobel Prize of Physics in 1957 to Chen Ning Yang and Tsung-Dao Lee for their insightful contributions to the matter.

\subsection{Mathematical aspects}

Chirality can be mathematically defined within any metric space by stating that an object is chiral if the group of isometries (metric preserving bijective transformations) under which it is invariant is generated by products of squared isometries \cite{Petitjean2010}. In the case of Euclidean space for instance, the  group of invariance must not contain any indirect isometries, namely centers of inversion, mirrors, and rotoreflection (or rotoinversion) axes. When an object is composite, the metric space and group of isometries to consider are of course the ones built from the cartesian product of the metric spaces and groups of isometries associated to each component of the object. The concept is thus considerably generalized, though it can lead to confusion if the underlying space is not explicitly specified. A molecule for instance might be geometrically achiral, but associated with a chiral graph (the space isometries then are graph automorphisms, which all must decompose into even numbers of edge preserving vertex transpositions) \footnote{A graph is a pair $G = (V,E)$ of a set $V$ of vertices and a set $E$ of edges such that $E \subseteq V \times V$ (which means that every edge joins either two vertices or a vertex to itself). A graph automorphism is a permutation $\pi$ of the set of vertices $V$ such that the pair of vertices $(v_1, v_2)$ forms an edge if and only if the pair $(\pi(v_1),\pi( v_2))$ of transformed  vertices forms en edge too. These edge-preserving transformations make up the automorphism group of a graph. A graph automorphism is a product of squared graph automorphisms if and only if it decomposes into an even number of edge-preserving transpositions. It follows from the general definition of chirality that if all the automorphisms of a graph are generated by products of even numbers of edge-preserving transpositions then the graph is chiral.}. As a whole it is chiral (this amounts to care not only about the geometry of the molecule but also about its constitutive elements as well as of the intramolecular interactions). Chirality depends on the space dimensionality. A chiral object in two dimensions, such as a non isosceles triangle, becomes achiral in three dimensions (see figure \ref{fig:1}), since the plane containing the 2-dimensional object becomes then a mirror symmetry. As a matter of fact, every $n$-dimensional object in a $d$-dimensional space is achiral as soon as $n<d$. It is indeed trivially identical to its (unchanged) image with respect to any $(d-1)$-dimensional hyperplane mirror containing the object.

\subsection{Crystallographic aspects}

An alternative definition is used in crystallography, which refers to the concept of centrosymmetry rather than mirror symmetry (we recall that a mirror is merely a combination of a two-fold rotation and a spatial inversion). It stipulates that a chiral object is not superposable by pure rotation and translation on its image formed by inversion through a point \cite{Flack2003}. We remind that an inversion transformation amounts to transforming every point to the point of opposite coordinates if its fixed point is taken as the space origin. It is then customary to call it parity and denote it with the symbol P. A necessary condition for a crystal structure to be chiral is that it is non-centrosymmetric. This is not sufficient however. The point group must contain only proper rotation symmetries. Non-centrosymmetric crystal structures can be achiral when the point group contains improper symmetry elements and when the structure and its image by inversion symmetry can be brought in coincidence by pure rotation. Centrosymmetric crystal structures on the other hand are always achiral. The two enantiomorphs of a chiral structure may belong either to the same space group (an example is the non-centrosymmetric space group $P321$) or to two distinct space groups transforming into each other by inversion. There are 11 pairs of such space groups, for instance $P6_1$ and $P6_5$, among the 65 space groups containing only proper symmetry elements, compatible with chiral structures. Optical activity had been historically an efficient mean to identify crystal chirality, but had been also misleading because there is no one-to-one correspondence between the two properties. Optical activity had been observed in achiral single crystals (belonging to the $m$, $mm2$, $\bar{4}$ and $\bar{4}2m$ crystal classes). Chirality of crystals can be adequately determined by anomalous (or resonant) X-ray scattering (or electron scattering), as first achieved by J.M. Bijvoet {\it et al.} on the tartaric acid in 1951 \cite{Bijvoet1951}. Owing to absorption induced phase shift, the scattering factor acquires an imaginary component and the scattering intensities at the opposite scattering vectors $\{(h,k,l), (-h,-k,-l)\}$ have no longer to be  necessarily equal. A systematic measurement of these Friedel pairs allows estimating the enantiomorph polarity of real crystals \cite{Flack1983}.

\subsection{Dynamical aspects}

Chirality becomes even more subtle when dynamical aspects have to be taken into account because one has to care about the time inversion. It is customary to denote this transformation with the symbol T. It is accounted for by an anti-unitary operator in quantum mechanics. Classically, it may be interpreted as reversing all the motions, effective and stationary. If a physical quantity is linearly built over a motion (linear momentum, angular momentum, magnetic moment $\equiv$ curl of a current density  field $\equiv$ infinitesimal current loop circulation, $\cdots$) then it will be reversed too, in which case it is said T-odd. If a physical quantity is invariant under T (linear translation, angular displacement, electric moment, $\cdots$) then it is said T-even. We remind that if a physical quantity is space dependent then it might be reversed under the space inversion P (linear translation, linear momentum, electric moment, $\cdots$), in which case it is said P-odd (or pseudo-scalar if it is a scalar and more generically polar). If a physical quantity is invariant under P (angular displacement, angular momentum, magnetic moment, $\cdots$) then it is said P-even (or true scalar if it is a scalar and more generically axial). A physical quantity most generally might be a mixed tensor (or a spinor), P-odd or P-even and T-odd or T-even in specified components. It is clear that the effects of the space inversion P and of the time inversion T coincide or differ according to the involved physical quantities, which might alter the concept of chirality. It matters in particular to wonder whether spatial enantiomorphism is still a sufficient condition of chirality. As an example, a collinear arrangement of an electric field, which is a T-even polar vector field, and of a magnetic field, which is a T-odd axial vector field, generates an enantiomorphism since parallel and antiparallel configurations are interconverted by space inversion P and are non superposable. These configurations however are also interconverted by time inversion T combined with a rotation by the angle $\pi$. It appears that such an arrangement of fields fails to induce any enantioselective process, for instance to bias a chemical reaction towards the production in excess of a given enantiomer. It thus becomes questionable to consider it as effectively chiral. According to Barron \cite{Barron1986}: "True chirality is exhibited by systems that exist in two distinct enantiomorphic states that are interconverted by space inversion but not by time reversal combined with any proper spatial rotation and translation". A spatial enantiomorphism that can be recovered by time-inversion combined with a proper spatial rotation or a translation is said associated with false chirality to emphasize the distinction. A true chirality is therefore distinguished from a false one if it is characterized by a T-even pseudo-scalar quantity. An example of true chirality is that of a coherent beam of photons with a given (either positive or negative) helicity/chirality (massless boson). One here is concerned with a T-odd polar vector (linear momentum) and a T-odd axial vector (photon spin), the scalar product of which is a T-even pseudo scalar (helicity $\equiv$ projection of the photon spin on its linear momentum). It has been demonstrated that (right- or left-) circularly polarized light can effectively influence the chirality of photochemical reaction products \cite{Avalos}.  

\section{Chirality in magnetism}
\label{MagnChirality}

\subsection{Influence of chirality in electromagnetic phenomena}

Chirality is often misapprehended in magnetism. A confusion for instance prevailed for long between the optical activity that occurs when a linearly polarized light crosses a media lacking mirror symmetry and the Faraday rotation that occurs when a linearly polarized light longitudinally crosses a media permeated by a static  magnetic field. As earlier emphasized by Lord Kelvin, quoting Michael Faraday who was aware of the fact, the two phenomena are basically distinct \cite{Kelvin1904}. When the light path is reversed the Faraday rotation changes sign whereas the optical activity does not. The mistake led to a series of misleading scenarios of absolute enantioselectivity (ability to produce an enantiomeric excess in an otherwise racemic mix), which have failed. Of course, this does not mean that a static magnetic field never contributes to an enantioselection. An example is magnetochiral dichroism \cite{Rikken1997}, which would arise from slightly different absorptions of light by chiral molecules according to whether the light beam, whatever its polarization, travels parallel or antiparallel to an externally applied magnetic field. A T-odd polar vector, the light wavevector, is here combined with a T-odd axial vector, the magnetic field, to induce a true chiral influence. One may furthermore emphasize that enantioselection by falsely chiral influence is {\it a priori} not excluded in case of irreversible processes and more generally in far from equilibrium systems \cite{Barron1986}.

\subsection{Vector chirality in spin networks}

In the context of spin networks, chirality is not a more evident concept. It covers, according to the relevant physical phenomena, basically distinct meanings, which are distinguished by specific names, vector chirality and scalar chirality, or may require more suitable explicit definitions. Among all the variants, the vector chirality is certainly the most intuitive, since it directly refers to the geometric image of a spin configuration, but at the same time it is subject to confusions and misconceptions. It is customary to conceive this chirality as a quantity that should indicate the sense of spin rotation when one is moving on oriented loops, typically on triangles or squares of spins, or along oriented lines, such as in helicoidal or cycloidal spin configurations. A generic variable answering this request is the spin twist or the chiral vector $$\vec\Xi_{ij} = \vec S_i \times \vec S_j$$ built over two consecutive spins $\vec S_i$ and $\vec S_j$ on an oriented link. According to the spin network and the spin configuration the spin twists must be possibly added over specific links, for instance over those of an oriented square, as the first time this chirality was invoked \cite{Villain1977}, or over those of an oriented triangle. The vector chirality associated with a triangle oriented by numbering its corners $(1, 2, 3)$ is thus generally given as $$\vec\Xi_\Delta = {2\over 3\sqrt{3}}{1\over \mathrm{S}^2} (\vec S_1 \times \vec S_2 + \vec S_2 \times \vec S_3 +\vec S_3 \times \vec S_1)$$ $\vec\Xi_\Delta$ is renormalized to get a variable that takes the values $\pm 1$ for a perfect triangular arrangement of the spins on the oriented triangle. It is clear that if the chirality is uniform then it is fully determined on a whole network through its average, either over all its oriented links or over all its oriented loops, otherwise one can have recourse to the Fourier Transform of the distribution of link or loop vector chiralities. One thus distinguishes for instance a ferrochiral spin arrangement (chiral vector identical on all loops), which is macroscopically characterized by a uniform vector chirality, from an antiferrochiral spin arrangement (chiral vector alternating from loop to loop), which is macroscopically characterized by a staggered vector chirality. In short, whatever the spin network and the spin configuration, a rigorous definition of the vector chirality can be formulated by making use of the spin twist variables $\vec\Xi_{ij}$. 

\begin{figure}
\resizebox{1.0\columnwidth}{!}{
\includegraphics{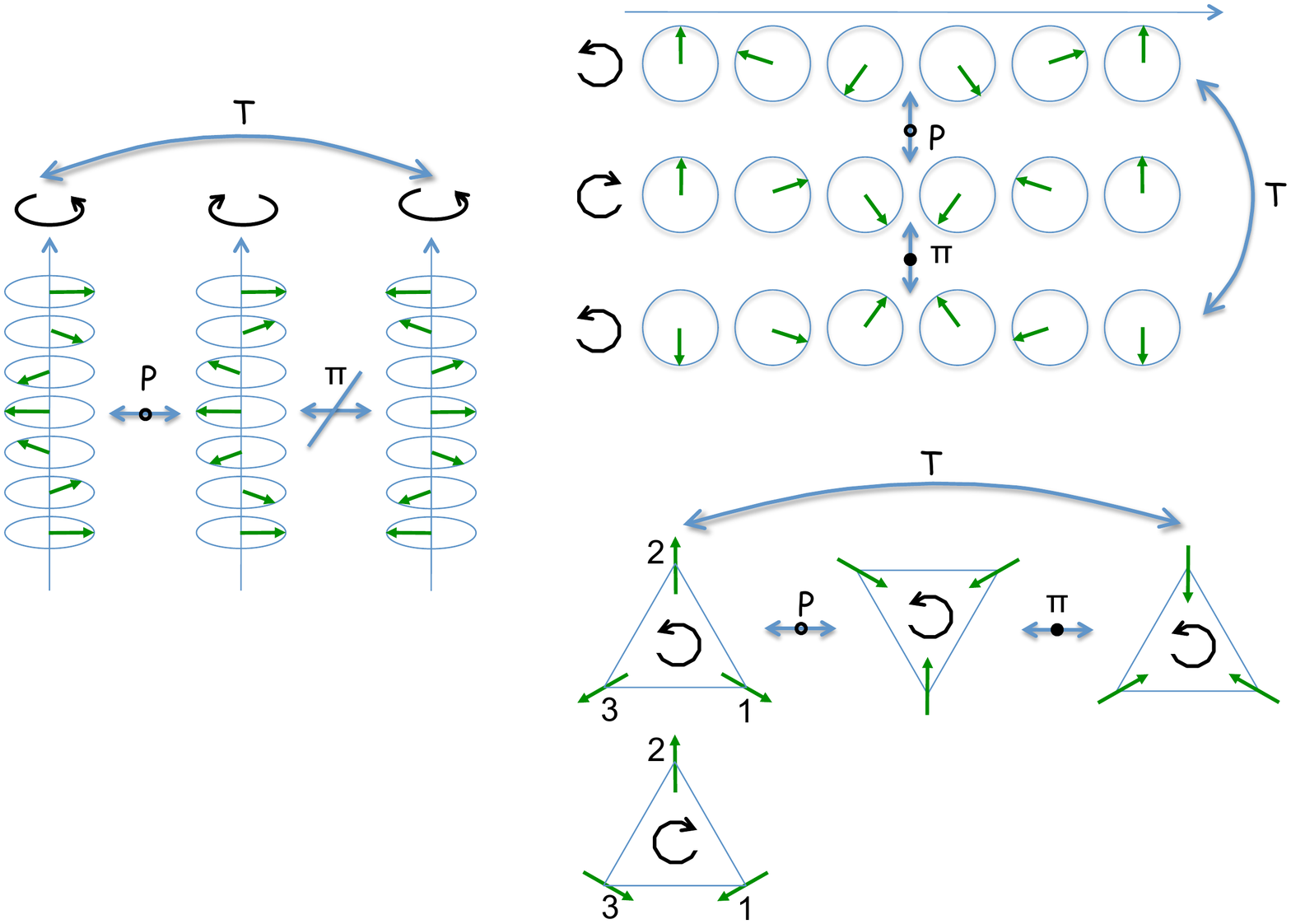} }
\caption{Examples of spin arrangements with magnetic helices (left), magnetic cycloids (top right), and moments at 120$^{\circ}$ on a triangle (bottom right). Their transformations through symmetry operations such as parity P (the center of inversion is represented by an empty dot), time-reversal T, and a two-fold axis $\pi$ (axis perpendicular to the figure plane intersected through the black dot for the triangle and cycloid) are shown in a geometric description. The spin currents are materialized by black curved arrows. In the case of magnetic helices, the enantiomorphic helix generated by P cannot be brought into coincidence with the initial helix after T and $\pi$ operations whatever the location and direction of the two-fold axis. }
\label{fig:2}       
\end{figure}

\subsubsection{Inconsistencies with naive geometric description}

The concept of vector chirality does not always meet the intuition of chirality we might forge from a purely geometric description of the spin configurations and their transformations by the ordinary space isometries. As to illustrate this, let us consider three typically encountered spin configurations showing a vector chirality:
 
- A magnetic helix, constituted of coplanar spins that rotate periodically about the perpendicular axis as one moves along it. One easily discerns two enantiomorphs interconverted by space inversion P. They are effectively associated with opposite vector chiralities, since the spins are rotating in opposite senses around the helix axis. The effect of P is impossible to produce with the time inversion T combined with a proper spatial rotation (see figure \ref{fig:2}). According to Barron, this would describe a true chirality. The concept of vector chirality in this case fits with our geometric intuition.

- A magnetic cycloid, constituted of coplanar spins that rotate periodically about the perpendicular axis as one moves along a specified axis parallel to the spin plane. One again discerns two enantiomorphs interconverted by space inversion P to which opposite vector chiralities are associated, but now these are also interconverted by the time inversion T combined with a 2-fold rotation. According to Barron, this would describe a false chirality (see figure \ref{fig:2}). A magnetic cycloid moreover is a $2$-dimensional object which should be achiral in 3 dimensions. Yet, one is still able to  unambiguously assign a distinct vector chirality to each of the enantiomorphs. 
 
- A triangular configuration of coplanar spins oriented at 120$^{\circ}$ from each other and distributed on the numbered corners of an equilateral triangle. Under the space inversion P the triangle is reversed, but not the spins which are axial. One gets a distinct enantiomorph, but which again can be recovered by the time inversion T combined with a 2-fold rotation (see figure \ref{fig:2}) and thus would describe a false chirality. What is worse now is that the vector chirality is identical for both enantiomorphs. Opposite vector chiralities are obtained solely by exchanging two spins out of three, which would suggest to consider the triangle not only as a geometric object but also as a graph. It is finally emphasized that once more the system should no longer be chiral in 3 dimensions and yet one is able to unambiguously distinguish the truly chiral spin configurations on the equilateral triangle with the help of the vector chirality $\vec\Xi_\Delta$.

It is obvious from the above three examples that the concept of vector chirality does not always match with the immediate perception we might have of the spin configurations and with the definition of chirality discussed in section \ref{intro}. As a matter of fact, the chiral vector cannot describe a true chirality according to Barron since it is not a pseudo-scalar. An helix however displays a true chirality because it can be characterized by the time-even pseudo scalar invariant $\vec S\cdot(\nabla\wedge\vec S)$ \cite{Dzyaloshinsky1964}. In the case of the cycloid, this quantity is zero and the only invariant that can be formed is of the type $(\vec S\cdot\nabla)\vec S-(\nabla\cdot \vec S)\vec S$ which is of the same nature as the vector chirality. With the triangular configuration of spins on a triangle, neither the true nor the false chirality concepts of Barron works since the parity does not connect the enantiomorphs of opposite vector chirality.

\subsubsection{Spin current and vector chirality}

What the inconsistencies of the naive description tells us is that the concept of vector chirality does not necessarily refer to the geometric inversion P, but to a more abstract transformation Q  that reverses the chiral vector from one enantiomorph to the other and forms with the identity Q$^2$ a group $Z_{2\rm Q} =\{\rm Q, \rm Q^2\}$ isomorphic to $Z_2 = \{-1, +1\}$, similarly as $Z_{2\rm P} = \{\rm P, \rm P^2\}$. The transformation Q differs from the parity P in a triangle of spins, which leads to a distinct Q$-$chirality (see figure \ref{fig:2}). The two transformations are identical in their effects on the spin configurations, for the magnetic helix and cycloid where it leads to respectively true and false chirality according to Barron. In these cases the Q$-$chirality coincide with P$-$chirality. 

The concept of vector chirality would loose any interest if it was only associated with an abstract transformation. It is actually supported by the fact that it can be given a concrete and intuitive meaning. A spin indeed is a physical quantity subject to dynamics and fluctuations driven by the interactions it experiences. With (nearest neighbor) isotropic bilinear exchange interactions $\sum_{\left<i,j\right>} \mathrm{J}_{ij} ~\vec S_i^\mathrm{o} \cdot \vec S_j^\mathrm{o}$ the equation of motion for the spin operator $\vec S_i^\mathrm{o}$ at the position $i$ writes $$\hbar {{\partial \vec S_i^\mathrm{o}} \over {\partial t}} + \sum_j \mathrm{J}_{ij} ~(\vec S_i^\mathrm{o} \times \vec S_j^\mathrm{o}) = 0$$ which is nothing but the discretized form of the continuity equation for spin conservation, by interpreting the quantity $\mathrm{J}_{ij} ~(\vec S_i^\mathrm{o} \times \vec S_j^\mathrm{o}) = \mathrm{J}_{ij} ~\vec \Xi_{ij}^\mathrm{o}$ as a {\bf spin current} operator from the site $i$ to the site $j$ \footnote{\label{spincurrent}It has been argued that, for localized spins, it is only the spin current operator component $\mathrm{J}_{ij} ~\vec \Xi_{ij}^\mathrm{o} -  \mathrm{J}_{ij} ~(\vec \Gamma_{ij} \cdot \vec \Xi_{ij}^\mathrm{o}) ~\vec \Gamma_{ij}$ perpendicular to the unit vector $\vec \Gamma_{ij} = (\left<\vec S_i^\mathrm{o}\right> \times \left<\vec S_j^\mathrm{o}\right>) / |\left<\vec S_i^\mathrm{o}\right> \times \left<\vec S_j^\mathrm{o}\right>|$ that effectively gives rise to an actual magnetization transport. It vanishes in the classical ground state or within a mean field approximation but not for quantum correlation
 \cite{Schutz2004}.}. Now it is clear that what makes up the vector chirality of a spin configuration is the spin current. It is observed that the spin current is intrinsically a $3-$dimensional quantity, since it arises from a pair of necessarily non-collinear spins and is aligned along a direction that makes up a frame with the two spins from which it is built. It thus always defines a $3-$dimensional chiral object whatever the dimension of the underlying network. Although the Q$-$chirality associated with the spin current $\mathrm{J}_{ij} ~\vec \Xi_{ij}^\mathrm{o}$ can no longer be apprehended according to the Barron's classification of true and false P$-$chirality, the dynamical aspects are still relevant since included in the equation of motion for the spins. The spin current is always T-even (polar T-even as a graph variable and axial T-even as a vector product of two spin variables). The Q$-$enantiomorphism in short should not be regarded as geometric but physical including the spin dynamics which is simultaneously time reversed. 

\subsubsection{Physics of spin vector chirality}

A series of microscopic mechanisms exist that might give rise to spin configurations displaying a finite vector chirality. A magnetic helix for instance can be induced from a ferromagnetic order by the Dzyaloshinskii-Moriya (DM) antisymmetric interactions \cite{Dzyaloshinsky1964}, in which case one generally expects that the pitch vector, which scales with the ratio of the DM coupling strength over the spin stiffness, should be of small amplitude, that is to say the spin configuration should show a long period. Antisymmetric interactions such as the DM ones actually might induce a multitude of spin textures, cycloidal, conical, $\cdots$, or else skyrmionic \cite{Muhlbauer2009,Pappas2009}. A magnetic helix might also be stabilized directly from the paramagnetic phase as an outcome of a competition between bilinear spin-spin exchange interactions \cite{Yoshimori1959} or merely owing to a structural twist of exchange paths \cite{Marty2008}. An ingredient of utmost relevance is geometric frustration, which is at the origin of the triangular spin configuration on a triangle of spins with antiferromagnetic nearest neighbor interactions and gives rise for instance to either the uniform or the staggered vector chirality in the kagom\'e network \cite{Grohol2005}. It may lead to an extremely wide variety of collective spin states and excitations in an extended network, depending on the network connectivity and numerous second order interactions and mechanisms, more or less susceptible to raise the frustration induced macroscopic degeneracies \cite{Ballou1998}. Usually the encountered magnetic structures show domains of opposite vector chiralities in a same proportion so that no net vector chirality emerges macroscopically. Magnetic chirality was for long observed as single domain only in the non-centrosymmetric itinerant helical magnet MnSi \cite{Ishida85}. A few other examples of non-centrosymmetric magnets exhibiting single domain vector chirality were later found out \cite{Marty2008,Janoshek2010}. It recently was brought to light that the vector chirality might play an important role in the field of multiferroism, where the spin current was identified as an essential ingredient of a possible mechanism of magneto-electric coupling \cite{Katsura2005}. In this description, the electric polarization produced between two spins $\vec S_i$ and $\vec S_j$ in the crystal is given by $\vec P_{ij}\propto \vec e_{ij}\times (\vec S_i\times \vec S_j)$ where $\vec e_{ij}$ stands for the unit vector connecting the two spin positions. This opens opportunities to handle magnetic chirality/ferroelectric domains by means of electric and/or magnetic fields. As a matter of fact, this had been experimentally demonstrated before \cite{Siratori1980}. Also promising is the electric handling of spin waves as a proposed route towards magnonics, a new way to carry and process information \cite{Rovillain2010}. This technological use of spin chirality is further highlighted by the present interest in the chiral character of magnetic objects in reduced dimensions (surfaces, magnetic domain walls...) \cite{Bode2007}. Chirality can also emanate from the quantum nature of the spin ensemble. An example is the soliton excitations in spin $S={1 \over 2}$ anisotropic chains \cite{Braun2005}, which has to do with the fact that the rotation group in spinor space is the double cover of the rotation group in ordinary space (more intuitively a spin $S={1 \over 2}$ is invariant under a 4$\pi-$rotation but is not under a 2$\pi-$rotation). On can also mention the prediction of critical behaviors associated with a universality class specific to chirality \cite{Kawamura1998} that are actively investigated. At last but not the least, a coupling of chirality vectors might occur on its own to possibly induce novel spin phases, for instance spin gels associated with the binding of vortices formed of chirality vectors \cite{Okubo2010}. All in all, a wealth of phenomena is inherent to the concept of the vector chirality, making up an extensive list which is constantly updated by new findings. 

\subsection{Scalar chirality in spin networks}

Chirality becomes further less intuitive when one is dealing with non coplanar magnetic structures all the more as the underlying network of spins cannot itself be approximated in terms of weakly coupled chains or planes of spins. A quantity often invoked for non coplanar spins is scalar chirality, which is defined for consecutive spins along an oriented line or on an oriented loop as $$\chi_{ijk} = \vec S_i \cdot (\vec S_j\times \vec S_k) $$ It is observed that this quantity, unlike the vector chirality, though still polar T-even as a graph variable, is now axial T-odd as the mixed product of three spin variables. Scalar chirality is rather relevant of (PT)$-$invariant physics, of which one of the exotic paradigm is provided by the anyon ensembles \cite{Khare2005}, which thus belong to the world of false chirality. It is inherent to the anomalous Hall effect observed in geometrically frustrated magnets \cite{Taguchi2001}. It allows characterizing the magnetic nature of multi-spin order phases with zero on-site spin average and predicting possible associated non trivial charge dynamics \cite{Khomskii2010}. It is an essential ingredient of many avatars of chiral spin liquids bearing analogy with liquid crystals \cite{Onoda2007}. Scalar chirality can be interpreted to some extent as a measure for the solid angle between three spins, but can also be given other equivalent meanings \cite{Wen1989}. One of these is in terms of spin $S={1 \over 2}$ circulations and associated fluxes, through the spin transfer bond operators $\phi_{ij} = c^+_{i\sigma}c_{j\sigma}$ where $c^+_{i\sigma}$ ($c_{j\sigma}$) is meant for the creation (annihilation) of a spin state $\sigma$ on the network node $i$ ($j$): $\chi_{123} = \left<\vec S_1^\mathrm{o} \cdot (\vec S_2^\mathrm{o}\times \vec S_3^\mathrm{o})\right> =  2 i \left<\phi_{12}\phi_{23}\phi_{31}\right> - \left<\phi_{13}\phi_{32}\phi_{21}\right>$ for a triangle numbered $(1, 2, 3)$. Another  invoked concept is the Berry's phase $B_{123} = \left<\mathrm{P}_{(123)}\right> = \left<\mathrm{P}_{(12)}\mathrm{P}_{(23)}\right> =  \left<(1 + \vec S_1^\mathrm{o} \cdot \vec S_2^\mathrm{o})(1 + \vec S_2^\mathrm{o} \cdot \vec S_3^\mathrm{o})\right> / 4$ associated with the cyclic transport of spin $S={1 \over 2}$ around a triangle numbered $(1, 2, 3)$ ($\mathrm{P}_{(ijk)}$ stands for the cyclic permutation operator and $\mathrm{P}_{(ij)}$ for a transposition): $\chi_{123} = -2 i (B_{123} - B_{132})$.

\vskip 0.2 cm

Although further aspects of the scalar chirality and other concepts of magnetic chirality could be evoked we shall here stop this brief overview of chirality in magnetism, which was intended to precise its physical meanings and to give a glimpse at the wealth of phenomena accounted for by this concept. We shall now describe one technique specially well suited to probe it, more precisely the vector chirality. 

\section{The Neutron Probe}
\label{neutron}	

A moving neutron is a true chiral object, by its linear momentum which is a T-odd polar vector and its spin which is a T-odd axial vector. It is thus suited to probe the magnetic chiralities in the spin networks. One in fact makes use of a beam of neutrons in the actual experiments, that is to say a statistical ensemble, so that one handles a beam polarization rather than the spinor of individual moving neutrons. It is obvious that only a polarization dependent contribution to the neutron scattering might probe a magnetic chirality. It is shown that such a contribution exists indeed, which we shall call chiral scattering. It however does not provide with an immediate access to the vector chirality. In addition the neutron allows probing primarily pair correlation functions. Chirality correlations, involving at least four spins, can be detected only indirectly. In order to specify these points more concretely we shall call back in what follows the general formulae of the intensity and polarization of the neutrons scattered out of a magnetic material, set up independently by Blume and Maleyev \cite{Blume1963,Maleyev1963}. We shall next briefly describe the methods of longitudinal and spherical neutron polarimetry to measure the chiral scattering \cite{Regnault2005} and indicate to what extent this might be significant of static and possibly dynamic magnetic chirality. 

\subsection{Blume-Maleyev Equations}

A targeted objective is the absolute determination of the chiral vectors. Unfortunately, it appears that there is no universal convention for the definition of the scattering quantities on which a chirality might depend. According to the context, crystallography, x-ray scattering, elastic or inelastic neutron scattering, $\cdots$, the plane waves may correspond to opposite wavevectors, the scattering vectors may differ in sign, $\cdots$, the geometry of the scattering is not always explicit, $\cdots$, which might lead to mistakes on the sign of the inferred chirality. We shall therefore begin with the basic concepts of the scattering theory, using the formulations implicitly adopted by Blume \cite{Blume1963}:

An incoming neutron plane wave $|~\vec k_i ~\sigma_i >$ propagating along a wavevector $\vec k_i$ is represented in space and time as 
$$\left< \vec r~t ~|~\vec k_i ~\sigma_i \right> = e^{i(\vec k_i \vec r-\omega_i t)} \quad \mathrm{with~} \omega_i = \hbar \vec k_i^2 / 2m$$

An outgoing neutron scattered wave is asymptotically represented in space and time, that is to say far from the scatterer, as 
$$\psi_{scat} = e^{i(\vec k_i \vec r-\omega_i t)} +  f(\vec k_i \sigma_i, \vec k_f \sigma_f) {e^{i(k_fr-\omega_f t)}\over r}  \quad \mathrm{with~} \omega_f = \hbar \vec k_f^2 / 2m$$ 
where $f(\vec k_i \sigma_i, \vec k_f \sigma_f)$ is the scattering amplitude. It follows that, for ingoing neutrons all in the state $|~\vec k_i ~\sigma_i >$ with kinetic energy $\hbar\omega_i$, the number of neutrons in the spin state $\sigma_f$ scattered per unit time within an infinitesimal solid angle $d\Omega$ along the unit vector $\vec{k}_f / k_f$  at a kinetic energy within an infinitesimal range $dE_f$ around $\hbar\omega_f$ is
\footnote{A probability density $P = \psi^*\psi$ and a probability current $\vec J = {\hbar \over 2m i} (\psi^* (\nabla \psi) - (\nabla \psi)^* \psi)$ are associated with every wavefunction $\psi$. It follows that $P = 1$ and  $\vec J = {\hbar \vec k \over m}$ for a plane wave $e^{i(\vec k \vec r-\omega t)}$ whereas $P = A^* A / r^2$ and $\vec J = {\hbar \vec k \over m} A^* A / r^2$ for a spherical wave $A e^{i(\vec k \vec r-\omega t)}/ r$.}
$$d^2\mathcal{N} = (\hbar k_f / m) ~f(\vec k_i \sigma_i, \vec k_f \sigma_f)^* f(\vec k_i \sigma_i, \vec k_f \sigma_f) ~\delta(\hbar\omega_f - \hbar\omega_i + E_{\lambda_f} - E_{\lambda_i}) ~d\Omega ~dE_f$$ 
for a collision process by which the state of the scatterer is transformed from $|~\lambda_i>$ with energy $E_{\lambda_i}$ to $|~\lambda_f>$ with energy $ E_{\lambda_f}$. The delta function $\delta(\hbar\omega_f - \hbar\omega_i + E_{\lambda_f} - E_{\lambda_i})$ ensures the energy conservation. 
Dividing by the flux $ (\hbar k_i / m)$ of the incoming neutrons one obtains a cross-section $d^2\sigma = (\partial^2\sigma / \partial\Omega \partial E_f) ~d\Omega ~dE_f$. Statistically averaging over the initial states  $|~\lambda_i>$ of the scatterer and the spin states $|~\sigma_i>$ of the incoming neutrons and summing over all the final states $|~\lambda_f>$ of the scatterer and the spin states $|~\sigma_f>$ of the outgoing neutrons, the partial differential cross-section is written
$$\left({\partial^2\sigma\over \partial\Omega \partial E_f}\right) = {k_f\over k_i} \sum_{\lambda_i, \sigma_i, \lambda_f, \sigma_f} p_{\lambda_i} ~p_{\sigma_i} ~|~f(\vec k_i \sigma_i, \vec k_f \sigma_f)~|^2 ~\delta(\hbar\omega_f - \hbar\omega_i + E_{\lambda_f} - E_{\lambda_i})$$
The statistical weight $p_{\lambda_i}$ depends on the scatterer. At equilibrium it may for instance be the Boltzmann factor:  $$p_{\lambda_i} = {\exp\{-E_{\lambda_i} / k_B T\} \over \sum_{E_{\lambda_i}} \exp\{-E_{\lambda_i} / k_B T\}} $$
The statistical weight $p_{\sigma_i}$ depends on the incoming neutron beam polarization $\vec P_i$: any operator $\mathrm{O}^\mathrm{o}$ in the spin space of an incoming neutron may be written as a linear combination $\mathrm{O}^\mathrm{o} = u 1^\mathrm{o} + \vec v \cdot \vec \Pi^\mathrm{o}$ of the unit operator $1^\mathrm{o}$ and the Pauli operators $\Pi^\mathrm{o}_x, \Pi^\mathrm{o}_y, \Pi^\mathrm{o}_z$. If we do so for the density operator $\rho^\mathrm{o} = \sum_i |~\sigma_i> p_{\sigma_i} <\sigma_i~|$ of the statistical ensemble associated with the incoming neutron beam then we find that $Tr [\rho^\mathrm{o}] = \sum_i p_{\sigma_i} = 1 = 2 u$ and $Tr [\rho^\mathrm{o}  \vec \Pi^\mathrm{o}] = \vec P_i$ (ensemble polarization) $= 2 \vec v$, that is to say 
$$\sum_i |~\sigma_i> p_{\sigma_i} <\sigma_i~| = {1 \over 2}(1^\mathrm{o} + \vec P_i \cdot \vec \Pi^\mathrm{o})$$ 
In the Born approximation to the Lippmann-Schwinger equation for the scattering 
\begin{eqnarray}
\nonumber f(\vec k_i \sigma_i, \vec k_f \sigma_f) & = & -{m\over 2\pi\hbar^2} \left<\vec k_f ~\sigma_f~\lambda_f ~|~\mathrm{V}^\mathrm{o}(\vec r)~| ~\vec k_i ~\sigma_i~\lambda_i\right> \\
\nonumber     & = & -{m\over 2\pi\hbar^2}\left<\sigma_f~\lambda_f ~|~\mathcal{V}^\mathrm{o}(\vec k_i - \vec k_f)~| ~\sigma_i~\lambda_i\right> 
\end{eqnarray}
where the operator $\mathrm{V}^\mathrm{o}(\vec r)$ stands for the interaction potential of the neutron with the scatterer and $\mathcal{V}^\mathrm{o}(\vec k_i - \vec k_f) = \int d\vec r ~ \mathrm{V}^\mathrm{o}(\vec r) e^{i (\vec k_i - \vec k_f) \cdot \vec r}$ is its Fourier transform. The wavevector 
$$\vec Q=\vec k_i-\vec k_f$$
is the scattering vector associated with the scattering channel $|~\vec k_i ~\sigma_i> \rightarrow |~\vec k_f ~\sigma_f>$. It is interpreted as the linear momentum transferred from the neutron to the crystal. The energy transferred from the neutron to the crystal in this scattering channel is 
$$\hbar\omega = \hbar\omega_i - \hbar\omega_f = {\hbar^2 \over {2 m }} (\vec k_i^2 - \vec k_f^2)$$ 
The dominating contributions to $\mathrm{V}^\mathrm{o}(\vec r)$ are:

- The neutron-nuclei nuclear interaction potential, which, out of resonance and for a nucleon at position $\vec R$ with spin $\vec\Upsilon$, is given as $\mathrm{V}_N^\mathrm{o}(\vec r) = (2\pi\hbar^2 / m) ~\mathrm{b}^\mathrm{o} \delta(\vec r- \vec R^\mathrm{o} )$ with $\mathrm{b}^\mathrm{o} = \tau~1^\mathrm{o} +  \upsilon~ \vec \Upsilon^\mathrm{o} \cdot \vec\Pi^\mathrm{o}$. It is also customary to define  scattering lengths $b^{(\pm)}$ associated with the nucleus-neutron total spin states $\Upsilon \pm 1/2$, in terms of which we may write $\tau = \{(\Upsilon + 1) b^{(+)} + \Upsilon b^{(-)}\} / (2\Upsilon +1)$ and $\upsilon = \{b^{(+)} - b^{(-)}\} / (2\Upsilon +1)$.

- The interaction potential  between the neutron magnetic moment $g_n |e|\hbar/ 2m$ $(g_n = -1.91348, |e|\hbar/ 2m = 5.05095~10^{-27} J T^{-1})$ and the electron spin and orbital current densities $\vec j^\mathrm{o} $ in the scatterer, which writes $\mathrm{V}_M^\mathrm{o}(\vec r) = - (g_n |e|\hbar/ 2m) ~\vec\Pi^\mathrm{o} \cdot \vec B^\mathrm{o} (\vec r)$ with $\vec B^\mathrm{o} (\vec r) = (\mu_0/ 4\pi)\int\vec j^\mathrm{o} (\vec s) \times [(\vec r - \vec s) / |\vec r - \vec s|^3] ~ d\vec s$ \cite{Halpern1939}. 

It follows that the scattering amplitude in general will be composed of a sum of a nuclear part and a magnetic part: 
$$f(\vec k_i \sigma_i, \vec k_f \sigma_f) = f_N(\vec k_i \sigma_i, \vec k_f \sigma_f) + f_M(\vec k_i \sigma_i, \vec k_f \sigma_f)$$ 
from which one interestingly should expect interference nuclear-magnetic scattering. In a crystal, that is to say in a network of scattering centers at positions $\vec R_{\nu n} = \vec r_\nu + \vec R_n$ where $\vec r_\nu$ stands for a position in the unit cell and $\vec R_n$ for a cell position, we have \cite{MarshallLovesey},
$$f_N(\vec k_i \sigma_i, \vec k_f \sigma_f)  = - \sum_{\nu n} \left<\sigma_f~\lambda_f ~|~\tau_{\nu n}~1^\mathrm{o} +  \upsilon_{\nu n}~ \vec \Upsilon_{\nu n}^\mathrm{o} \cdot \vec\Pi^\mathrm{o}~| ~\sigma_i~\lambda_i \right> ~e^{i \vec Q \cdot \vec R_{\nu n}^\mathrm{o}}$$
$$f_M(\vec k_i \sigma_i, \vec k_f \sigma_f)  = -p ~{1 \over \vec Q^2} \left<\lambda_f ~|~ \left(\vec Q \times \vec M^\mathrm{o}(\vec Q) \times \vec Q \right) ~|~\lambda_i \right> \cdot \left<\sigma_f ~|~ \vec\Pi^\mathrm{o}~| ~\sigma_i \right>$$
$\vec M^\mathrm{o} (\vec Q) = \int \vec M^\mathrm{o} (\vec r) e^{i \vec Q \cdot \vec r} ~d\vec r$ where $\vec M^\mathrm{o} (\vec r)$ is the magnetization density operator in units of Bohr magneton ($\mu_B$) and $p=2.696 ~fm/\mu_B$. In most of the cases $\vec M^\mathrm{o} (\vec Q) = \sum_{\nu} f_\nu(\vec Q)\sum_{n} \vec m_{\nu n}^\mathrm{o} ~e^{i \vec Q \cdot \vec R_{\nu n}^\mathrm{o}}$ where $f_\nu(\vec Q)$ is a magnetic form factor, which depends only on the nature of the scatterer centers, and $\vec m_{\nu n}^\mathrm{o} $ the magnetic moment operator associated with the scatterer at position $\vec R_{\nu n}$ \cite{Ballou2005}. 

Now, whatever the operator $\mathrm{O}^\mathrm{o} = u 1^\mathrm{o} + \vec s \cdot \vec \Pi^\mathrm{o}$, it is established that 
\footnote{\label{footnotePauli}A useful identity is $\Pi_{\alpha}\Pi_{\beta} = \delta_{\alpha\beta} 1^\mathrm{o} + i \epsilon_{\alpha\beta\gamma}\Pi_{\gamma}$, where $\delta_{\alpha\beta}$ is the Kronecker symbol and $\epsilon_{\alpha\beta\gamma}$ the Levi-Civita symbol. It follows that $Tr[\Pi_{\alpha}\Pi_{\beta}] = 2 \delta_{\alpha\beta}$  and $Tr[\Pi_{\alpha}\Pi_{\beta}\Pi_{\gamma}] = 2i \epsilon_{\alpha\beta\gamma}$.}
$$\sum_{\sigma_i, \sigma_f} p_{\sigma_i} \left<\sigma_i~|~(\mathrm{O}^\mathrm{o})^\dag ~|~ \sigma_f \right> \left<\sigma_f~|~\mathrm{O}^\mathrm{o} ~|~ \sigma_i \right> = Tr[\rho^\mathrm{o} (\mathrm{O}^\mathrm{o})^\dag \mathrm{O}^\mathrm{o}]$$ $$ = u^* u + \vec s^* \cdot \vec s + u^* (\vec P_i \cdot \vec s) + (\vec P_i \cdot \vec s^*) u + i \vec P_i \cdot (\vec s^* \times \vec s)$$
Note that $u$ and $\vec s$ might be operators that act in a state space other than the neutron spin state space, in which case $u, \vec s, u^*, \vec s^*$ are merely replaced by $u^\mathrm{o}, \vec s^\mathrm{o}, (u^\mathrm{o})^\dag, (\vec s^\mathrm{o})^\dag$.
The nuclear interaction operator is of the type $u 1^\mathrm{o} + \vec v \cdot \vec \Pi^\mathrm{o}$ and the magnetic interaction operator of the type $\vec w \cdot \vec \Pi^\mathrm{o}$. So the average over the spin states $|~\sigma_i >$ of the incoming neutrons and the summation over all the spin states $|~\sigma_f >$ of the outgoing neutrons should lead to the equation $u^* u + \vec v^* \cdot \vec v + u^* (\vec P_i \cdot \vec v) + (\vec P_i \cdot \vec v^*) u + i \vec P_i \cdot (\vec v^* \times \vec v) + \vec v^* \cdot \vec w + \vec v \cdot \vec w^* + u^* (\vec P_i \cdot \vec w) + (\vec P_i \cdot \vec w^*) u + i \vec P_i \cdot (\vec v^* \times \vec w) + i \vec P_i \cdot (\vec w^* \times \vec v) + \vec w^* \cdot \vec w + i \vec P_i \cdot (\vec w^* \times \vec w)$. Assuming that the nuclear spins of the scatterer are randomly oriented, the averaging over the nuclear spin states cancels all the terms that are linear in the nuclear spin operators $\vec \Upsilon^\mathrm{o}_{\nu n}$, which leaves with $u^* u + \vec v^* \cdot \vec v +  u^* (\vec P_i \cdot \vec w) + (\vec P_i \cdot \vec w^*) u + \vec w^* \cdot \vec w + i \vec P_i \cdot (\vec w^* \times \vec w)$. 

Next, if $\Sigma^\mathrm{o}$ is an operator that acts only on the scatterer states then one may write 
\footnote{It suffices to use the identity $\delta(E_{\lambda_f}-E_{\lambda_i}-\hbar\omega) = {1\over 2\pi\hbar}\int_{-\infty}^{+\infty}e^{i(E_{\lambda_f}-E_{\lambda_i}){t\over h}}e^{-i\omega t}dt$ to write ${1\over 2\pi\hbar}\int_{-\infty}^{+\infty} \sum_{\lambda_i, \lambda_f} p_{\lambda_i} \left<\lambda_i |~(\Sigma^\mathrm{o})^\dag~| \lambda_f \right > \left<\lambda_f |~e^{i \mathcal{H} t / \hbar} \Sigma^\mathrm{o}e^{-i \mathcal{H} t / \hbar}~| \lambda_i \right > e^{-i\omega t}dt$ and then to use merely  the closure relation in the space of the scatterer states: $\sum_{\lambda_f} |\lambda_f > <\lambda_f |= 1^\mathrm{o}$.}
$$\sum_{\lambda_i, \lambda_f} p_{\lambda_i} |\left<\lambda_f |~\Sigma^\mathrm{o}~| \lambda_i \right >|^2 ~\delta(E_{\lambda_f}-E_{\lambda_i}-\hbar\omega) = {1\over 2\pi\hbar}\int_{-\infty}^{+\infty} < (\Sigma^\mathrm{o})^\dag \Sigma^\mathrm{o}(t) >e^{-i\omega t}dt$$
with $\Sigma^\mathrm{o}(t) = e^{i \mathcal{H}^\mathrm{o} t / \hbar} \Sigma^\mathrm{o}e^{-i \mathcal{H}^\mathrm{o} t / \hbar}$ where $\mathcal{H}^\mathrm{o}$ is the Hamiltonian of the scatterer and $<\cdots> = \sum_\lambda p_\lambda \left<\lambda | \cdots | \lambda \right>$ symbolizes the statistical average over the scatterer states. 

By putting $$\mathrm{N}^\mathrm{o}(\vec Q,t) = \sum_{\nu n} \tau_{\nu n}~e^{i\vec Q.\vec R_{\nu n}^\mathrm{o}(t)} \mathrm{~and~} \vec T^\mathrm{o}(\vec Q,t) = \sum_{\nu n} \upsilon_{\nu n} \vec \Upsilon_{\nu n}^\mathrm{o}(t)~e^{i\vec Q.\vec R_{\nu n}^\mathrm{o}(t)}$$ for the scattering by the nuclei and $$\vec M_{\perp}^\mathrm{o}(\vec Q,t) = p\sum_{\nu} f_\nu(\vec Q)\sum_{n} {1\over \vec Q^2}(\vec Q \times \vec m_{\nu n}^\mathrm{o}(t) \times \vec Q) ~e^{i \vec Q \cdot \vec R_{\nu n}^\mathrm{o}(t)}$$ for the scattering by the electron spin and orbital currents, one finally obtains the (more intuitive) formula:
$${d^2\sigma\over d\Omega dE_f}  =  {k_f\over k_i}\left({1\over 2\pi\hbar^2}\right) \int_{-\infty}^{+\infty}\Big\{<{\mathrm{N}^\mathrm{o}}^{\dag}(\vec Q,0)\mathrm{N}^\mathrm{o}(\vec Q,t)> +<\vec {T^\mathrm{o}}^{\dag}(\vec Q,0) \cdot \vec T^\mathrm{o} (\vec Q, t)>$$
\begin{equation}
\label{equ:crosssection}
+ \vec P_i \cdot <{\mathrm{N}^\mathrm{o}}^{\dag}(\vec Q,0)\vec M^\mathrm{o}_{\perp}(\vec Q,t)+\vec M^{\mathrm{o}\dag}_{\perp}(\vec Q,0)\mathrm{N}^\mathrm{o}(\vec Q,t)>
\end{equation}
$$+  <\vec M^{\mathrm{o}\dag}_{\perp}(\vec Q,0) \cdot \vec M^\mathrm{o}_{\perp}(\vec Q, t)> + i\vec P_i \cdot <\vec M^{\mathrm{o}\dag}_{\perp}(\vec Q,0)\times \vec M^{\mathrm{o}}_{\perp}(\vec Q,t)>\Big\} e^{-i\omega t}dt$$
The first two terms of the r.h.s. of the equation (\ref{equ:crosssection}) accounts for the nuclear scattering, assumed implicitly averaged over the isotopic distributions. The second two terms describe nuclear-magnetic interference in the scattering and the last two terms stand for the neutron scattering by symmetric and antisymmetric magnetic pair correlations. 

A similar computation can be performed to determine the beam polarization $\vec P_f$ of the scattered neutrons, which can also be measured. $\vec P_f$ is proportional to $Tr[\rho^\mathrm{o} (u 1^\mathrm{o} + \vec v \cdot \vec \Pi^\mathrm{o})^\dag ~\vec \Pi^\mathrm{o} ~(u 1^\mathrm{o} + \vec v \cdot \vec \Pi^\mathrm{o})]$ because it characterizes the transformation of the spin state of the incoming neutrons induced by the scattering process. Its modulus, by definition of a beam polarization, cannot range beyond the interval $[0,1]$ so that the constant of proportionality must be $(Tr[\rho^\mathrm{o} (u 1^\mathrm{o} + \vec v \cdot \vec \Pi^\mathrm{o})^\dag ~1^\mathrm{o} ~(u 1^\mathrm{o} + \vec v \cdot \vec \Pi^\mathrm{o})])^{-1}$, which is nothing but the inverse of the partial differential cross-section. It follows, using  the previous notations, that
\footnote{\label{scatpolequdetails}With the help of the identity $Tr[\Pi_{\alpha}\Pi_{\beta}\Pi_{\gamma}\Pi_{\eta}] = 2 (\delta_{\alpha\beta}\delta_{\gamma\eta}-\delta_{\alpha\gamma\beta}\delta_{\beta\eta}+\delta_{\alpha\eta}\delta_{\beta\gamma})$ in addition to the ones reported in the footnote \ref{footnotePauli}, one indeed easily shows that if $\mathrm{O}^\mathrm{o} = u 1^\mathrm{o} + \vec s \cdot \vec \Pi^\mathrm{o}$ then $Tr[\rho^\mathrm{o} (\mathrm{O}^\mathrm{o})^\dag \vec \Pi^\mathrm{o} \mathrm{O}^\mathrm{o}] = u^* \vec s + \vec s^* u + u^* u ~\vec P_i + \vec s^* (\vec s \cdot \vec P_i) + (\vec s^* \cdot \vec P_i) \vec s - \vec P_i ( \vec s^* \cdot  \vec s) - i \vec s^* \times \vec s + i u^* (\vec s \times \vec P_i ) + i ( \vec P_i \times \vec s^*) u$, where $u, \vec s, u^*$ and $\vec s^*$ must be merely replaced by $u^\mathrm{o}, \vec s^\mathrm{o}, (u^\mathrm{o})^\dag$ and $(\vec s^\mathrm{o})^\dag$ whenever instead of c$-$numbers we have to handle operators that act on a state space other than the neutron spin state space. An averaging over randomly oriented nuclear spin states cancels all the terms that are linear in the nuclear spins operator $\vec v =  \sum_{\nu n} \upsilon_{\nu n} \vec \Upsilon^\mathrm{o}_{\nu n}$, including $\vec v^* \times \vec v$, which leaves with $\vec P_i (u^* u) + \vec v^* (\vec P_i \cdot \vec v) + (\vec P_i \cdot \vec v^*) \vec v - \vec P_i ( \vec v^* \cdot  \vec v) + u^* \vec w + \vec w^* u +  i (\vec P_i \times \vec w^*) u + i u^* (\vec w \times \vec P_i) - \vec P_i ( \vec w^* \cdot  \vec w) + \vec w^* (\vec P_i \cdot \vec w) + (\vec P_i \cdot \vec w^*) \vec w - i \vec w^* \times \vec w$, by putting $\vec s = \vec v+\vec w$ namely by taking into account  the neutron-electron magnetic contribution $\vec w$. Note that the average over randomly oriented nuclear spin states of $\vec v^* (\vec v \cdot \vec P_i)$ and $(\vec v^* \cdot \vec P_i) \vec v$ should be the same number, equal to one third ($1/3$) of the similar average on $\vec P_i ( \vec v^* \cdot  \vec v)$.}
$$\vec P_f {d^2\sigma\over d\Omega dE_f} = $$
$$ {k_f\over k_i}\left({1\over 2\pi\hbar^2}\right) \int_{-\infty}^{+\infty}\Big\{\vec P_i <{\mathrm{N}^\mathrm{o}}^{\dag}(\vec Q,0)\mathrm{N}^\mathrm{o}(\vec Q,t> + <\vec {T^\mathrm{o}}^{\dag}(\vec Q,0) (\vec P_i \cdot \vec T^\mathrm{o} (\vec Q, t))>$$
$$ + <(\vec P_i \cdot \vec {T^\mathrm{o}}^{\dag}(\vec Q,0)) \vec T^\mathrm{o} (\vec Q, t)> - \vec P_i <\vec {T^\mathrm{o}}^{\dag}(\vec Q,0) \cdot \vec T^\mathrm{o} (\vec Q, t)>$$
$$ + <{\mathrm{N}^\mathrm{o}}^{\dag}(\vec Q,0)\vec M^\mathrm{o}_{\perp}(\vec Q,t)> + <\vec M^{\mathrm{o}\dag}_{\perp}(\vec Q,0)\mathrm{N}^\mathrm{o}(\vec Q,t)>$$
\begin{equation}
\label{equ:scatteredtpolarization}
+ i < (\vec P_i \times \vec M^{\mathrm{o}\dag}_{\perp}(\vec Q,0)) \mathrm{N}^\mathrm{o}(\vec Q,t)> + i <{\mathrm{N}^\mathrm{o}}^{\dag} (\vec P_i \times \vec M^\mathrm{o}(\vec Q,t))>
\end{equation}
$$ - \vec P_i <\vec M^{\mathrm{o}\dag}_{\perp}(\vec Q,0)\cdot \vec M^\mathrm{o}_{\perp}(\vec Q,t)> + <\vec M^{\mathrm{o}\dag}_{\perp}(\vec Q,0) (\vec P_i \cdot \vec M^\mathrm{o}_{\perp}(\vec Q,t)) >$$
$$+<(\vec P_i \cdot \vec M^{\mathrm{o}\dag}_{\perp}(\vec Q,0)) \vec M^\mathrm{o}_{\perp}(\vec Q,t) > - i < \vec M^{\mathrm{o}\dag}_{\perp}(\vec Q,0) \times \vec M^\mathrm{o}_{\perp}(\vec Q,t)>\Big\} e^{-i\omega t}dt $$
The first four terms of the r.h.s. of the equation (\ref{equ:scatteredtpolarization}) accounts for the effects of the nuclear scattering on the polarization. Considering the remark done at the end of the footnote \ref{scatpolequdetails} one finds that the beam polarization is unchanged by the coherent nuclear scattering and by the incoherent nuclear scattering due to the isotopic distribution but it is changed to $\vec P_f = -{1\over 3} \vec P_i$ by the incoherent nuclear scattering due to the random nuclear spin orientations. The following four terms in the equation (\ref{equ:scatteredtpolarization}) describe the polarization transformation induced by the nuclear-magnetic interferences in the scattering, among which the two first create a polarization from an initially unpolarized beam. The last four terms accounts for the effects of the magnetic scattering on the polarization, among which the last one arises from antisymmetric magnetic pair correlations. It creates polarization from an initially unpolarized beam. It is this antisymmetric magnetic scattering, we call {\bf chiral scattering}, which might give information about possible spin chirality in a spin network. We indicate in the following neutron techniques for its measurement. Note finally that additional terms depending on the cross product $\vec k_i\times \vec k_f$ between the ingoing and outgoing wavevectors should appear in expressions of the partial differential cross-section and the final polarization when relativistic and spin-orbit corrections are taken into account \cite{Schwinger1948,Blume1964}. A weak asymmetry would result in the scattering process. These terms are however expected to be three orders of magnitude smaller than the standard nuclear and magnetic ones.

\subsection{Longitudinal Polarimetry}\label{longitudinal}

The simplest neutron polarimetry technique to probe the chiral scattering was first introduced by Moon, Riste and Koehler \cite{Moon1969}, and is called longitudinal polarization analysis (LPA). In this case the final and initial polarizations are parallel, which can be achieved typically on a triple-axis spectrometer with polarizing monochromator/analyser ({\it e.g.} Heusler crystals). In addition, Helmoltz coils allow controlling the magnetic field direction, hence selecting the polarization direction, maintained by guiding fields along the neutrons path, whereas two flippers select the polarization states $+$ and $-$ (parallel or antiparallel to the polarization axis). The spin-flip terms (scattering processes changing the sign of the polarization), 
$$({d^2\sigma\over d\Omega dE_f})^{+-}=\sigma^{+-} \mathrm{~and~}({d^2\sigma\over d\Omega dE_f})^{-+}=\sigma^{-+}$$ 
and the non-spin-flip terms (scattering processes leaving the sign of the polarization unchanged), 
$$({d^2\sigma\over d\Omega dE_f})^{++}=\sigma^{++} \mathrm{~and~} ({d^2\sigma\over d\Omega dE_f})^{--}=\sigma^{--}$$ 
of the partial differential cross-section can be measured independently. 

\vskip 0.2 cm 

In neutron polarimetry, a right-handed coordinated system is usually chosen with the $x-$axis along the scattering vector $\vec Q$, the $y-$axis in the scattering plane and the $z-$axis perpendicular to the scattering plane, so that the magnetic interaction vector has zero $x$ component. We shall use in the following the simplified notations:

\begin{center}
\noindent$\sigma_N={k_f\over k_i}\left({1\over 2\pi\hbar^2}\right)\int_{-\infty}^{+\infty}<{\mathrm{N}^\mathrm{o}}^{\dag}(\vec Q,0)\mathrm{N}^\mathrm{o}(\vec Q,t)>e^{-i\omega t}dt$

\vskip 0.2 cm

\noindent$\sigma_M^y={k_f\over k_i}\left({1\over 2\pi\hbar^2}\right)\int_{-\infty}^{+\infty}<\vec M^{\mathrm{o}y\dag}_{\perp}(\vec Q,0) \vec M^{\mathrm{o}y}_{\perp}(\vec Q, t)>e^{-i\omega t}dt$

\vskip 0.2 cm

\noindent $\sigma_M^z={k_f\over k_i}\left({1\over 2\pi\hbar^2}\right)\int_{-\infty}^{+\infty}<\vec M^{\mathrm{o}z\dag}_{\perp}(\vec Q,0) \vec M^{\mathrm{o}z}_{\perp}(\vec Q, t)>e^{-i\omega t}dt$

\vskip 0.2 cm

\noindent$M_{ch}={k_f\over k_i}\left({1\over 2\pi\hbar^2}\right)\int_{-\infty}^{+\infty}i<\vec M^{\mathrm{o}\dag}_{\perp}(\vec Q,0)\times \vec M^\mathrm{o}_{\perp}(\vec Q,t)>^xe^{-i\omega t}dt$

$={k_f\over k_i}\left({1\over 2\pi\hbar^2}\right)\int_{-\infty}^{+\infty}i(<\vec M{^\mathrm{o}y\dag}_{\perp}(\vec Q,0) \vec M^{\mathrm{o}z}_{\perp}(\vec Q, t)>-<\vec M{^\mathrm{o}z\dag}_{\perp}(\vec Q,0) \vec M^{\mathrm{o}y}_{\perp}(\vec Q, t)>)e^{-i\omega t}dt$

\vskip 0.2 cm

\noindent$M_{yz}={k_f\over k_i}\left({1\over 2\pi\hbar^2}\right)\int_{-\infty}^{+\infty}(<\vec M^{\mathrm{o}y\dag}_{\perp}(\vec Q,0) \vec M^{\mathrm{o}z}_{\perp}(\vec Q, t)> + \qquad\qquad\qquad\qquad$

\vskip 0.2 cm

\noindent$\qquad\qquad\qquad\qquad + <\vec M^{\mathrm{o}z\dag}_{\perp}(\vec Q,0) \vec M^{\mathrm{o}y}_{\perp}(\vec Q, t)>)e^{-i\omega t}dt$

\vskip 0.2 cm

\noindent$R_y={k_f\over k_i}\left({1\over 2\pi\hbar^2}\right)\int_{-\infty}^{+\infty}(<{\mathrm{N}^\mathrm{o}}^{\dag}(\vec Q,0) \vec M^{\mathrm{o}y}_{\perp}(\vec Q, t)>+<\vec M^{\mathrm{o}y\dag}_{\perp}(\vec Q,0) \mathrm{N}^\mathrm{o}(\vec Q, t)>)e^{-i\omega t}dt$

\vskip 0.2 cm

\noindent$R_z={k_f\over k_i}\left({1\over 2\pi\hbar^2}\right)\int_{-\infty}^{+\infty}(<{\mathrm{N}^\mathrm{o}}^{\dag}(\vec Q,0) \vec M^{\mathrm{o}z}_{\perp}(\vec Q, t)>+<\vec M{^\mathrm{o}z\dag}_{\perp}(\vec Q,0) \mathrm{N}^\mathrm{o}(\vec Q, t)>)e^{-i\omega t}dt$

\vskip 0.2 cm

\noindent$I_y={k_f\over k_i}\left({1\over 2\pi\hbar^2}\right)\int_{-\infty}^{+\infty}(i<{\mathrm{N}^\mathrm{o}}^{\dag}(\vec Q,0) \vec M^{\mathrm{o}y}_{\perp}(\vec Q, t)>-<\vec M^{\mathrm{o}y\dag}_{\perp}(\vec Q,0) \mathrm{N}^\mathrm{o}(\vec Q, t)>)e^{-i\omega t}dt$

\vskip 0.2 cm

\noindent$I_z={k_f\over k_i}\left({1\over 2\pi\hbar^2}\right)\int_{-\infty}^{+\infty}(i<{\mathrm{N}^\mathrm{o}}^{\dag}(\vec Q,0) \vec M^{\mathrm{o}z}_{\perp}(\vec Q, t)>-<\vec M^{mathrm{o}z\dag}_{\perp}(\vec Q,0) \mathrm{N}^\mathrm{o}(\vec Q, t)>)e^{-i\omega t}dt$
\end{center}

The chiral scattering term $M_{ch}$ can be easily determined in longitudinal polarization analysis when the polarization is parallel to $x$, since

$$\sigma_x^{++}=\sigma_x^{--}=\sigma_N$$

$$\sigma_x^{+-}=\sigma_M^y+\sigma_M^z+P_iM_{ch}$$

$$\sigma_x^{-+}=\sigma_M^y+\sigma_M^z-P_iM_{ch}$$ so that

$${\sigma_x^{+-}-\sigma_x^{-+}\over 2P_i}=M_{ch}$$

\noindent This technique is free from parasitic nuclear and background contributions. 

Alternatively, one can use an initial unpolarized beam and perform polarization analysis. In this case the partial differential cross-section writes $\sigma_0=\sigma_N+\sigma_M^y+\sigma_M^z$ and one deduces the chiral scattering from the scattered neutron beam polarization as
$$M_{ch}=-\sigma_0P_f^x$$ 

The chiral scattering can also be derived from the difference:
$${\sigma_x^{0-}-\sigma_x^{0+}\over 2}=M_{ch}$$
where the symbol 0 is meant for zero beam polarization of the incoming neutrons.

A last method to get the chiral scattering consists in using a polarized beam of incoming neutrons and measuring the intensity of the scattered neutrons in all the polarization channels. The chiral scattering in this case is given as 
$${\sigma_x^{+0}-\sigma_x^{-0}\over 2P_i}=M_{ch}$$

\subsection{Spherical polarimetry}

The chiral scattering can also be determined using neutron polarimetry. This technique involves the CRYOPAD device \cite{Tasset1989}, acronym for Cryogenic Polarization Analysis Device. It is a zero-field chamber, where the sample is positioned, designed to choose independently the initial and final polarization directions, hence allowing to measure the non-diagonal elements of the polarization matrix $\mathbf{P}$. An element $P_{\alpha\beta}$ of $\mathbf{P}$ is understood as the polarization of the outgoing beam along the $\beta-$axis when the ingoing beam was polarized along the $\alpha-$axis. The zero magnetic field is achieved by superconducting and $\mu$-metal screens. A system of precession and nutator coils allows setting the polarization direction of the incoming beam and reorienting the selected polarization direction for the outgoing beam along a given quantization axis where the polarization amplitude is determined as $$P_f={I^+-I^-\over I^++I^-}$$ from the counts $I^+$ and $I^-$ of the neutrons in the  spin states $|+>$ and $|->$.

The polarization matrix, that can be derived from the Blume and Maleyev equations, is given by the following expression using the reduced notations:

$$\mathbf{P} = \left(
	\begin{array}{ccc}
P_{xx}  & P_{xy}  & P_{xz} \\
P_{yx}  & P_{yy}  & P_{yz} \\
P_{zx}  & P_{zy}  & P_{zz}
	\end{array}\right)$$
\begin{equation}
\label{Matricepolar}
 = \left(
	\begin{array}{ccc}
\frac{-M_{ch}+P_0(\sigma_N-\sigma_M^y-\sigma_M^z)}{\sigma_N+\sigma_M^y+\sigma_M^z+P_0M_{ch}}  & \frac{R_y+I_zP_0}{\sigma_N+\sigma_M^y+\sigma_M^z+P_0M_{ch}}& \frac{R_z-I_yP_0}{\sigma_N+\sigma_M^y+\sigma_M^z+P_0M_{ch}} \\
\frac{-M_{ch}-I_zP_0}{\sigma_N+\sigma_M^y+\sigma_M^z+P_0R_y} & \frac{R_y+P_0(\sigma_N+\sigma_M^y-\sigma_M^z)}{\sigma_N+\sigma_M^y+\sigma_M^z+P_0R_y} & \frac{R_z+P_0M_{yz}}{\sigma_N+\sigma_M^y+\sigma_M^z+P_0R_y} \\
\frac{-M_{ch}+I_yP_0}{\sigma_N+\sigma_M^y+\sigma_M^z+P_0R_z} & \frac{R_y+P_0M_{zy}}{\sigma_N+\sigma_M^y+\sigma_M^z+P_0R_z} & \frac{R_z+P_0(\sigma_N-\sigma_M^y+\sigma_M^z)}{\sigma_N+\sigma_M^y+\sigma_M^z+P_0R_z}
	\end{array}\right)
\end{equation}

The chiral scattering is present in the expression of the diagonal term $P_{xx}$. It contributes also to the $P_{yx}$ and $P_{zx}$ matrix components. Note that the non-diagonal matrix elements give access to the magnetic symmetric crossed correlation functions ($M_{yz}$) and to the antisymmetric nuclear-magnetic correlation functions ($I_y$ and $I_z$), that can not be determined by LPA. When the nuclear-magnetic interference contributions can be ignored, $P_{yx}$ and $P_{zx}$, which then are equal, provide directly the sign of the chiral term and then allow determining the absolute vector chirality of a magnetic structure. The chiral term can be fully determined from the sum of the  $P_{yx}$ (or $P_{zx}$) quantities obtained with a given initial polarization along $y$ ($z$) and the initial polarization with opposite sign:

$$M_{ch}=-{P_{yx}+P_{\bar yx}\over 2}\sigma_0=-{P_{zx}+P_{\bar zx}\over 2}\sigma_0$$

\subsection{Diffraction from Crystals}
\label{chiraldiff}

It is customary to distinguish among the  scattering processes of neutrons by a crystal \\ 
- the coherent scattering that arises from a perfectly periodic interaction potential, which in concrete materials is merely the actual potential averaged over all the static and dynamic crystal defects (it is worth emphasizing here, although this might go without saying, that collective excitations such as phonons, magnons, $\cdots$ are evidently not crystal defects but dynamics inherent to a perfect crystal), \\
- the intrinsic incoherent scattering due to random isotope distributions and random nuclear spin orientations, \\
- the diffuse scattering associated with all other deviations of the interaction potential from the one of a perfect crystal  (inherent for instance to nuclear or magnetic impurities, interstitial or vacancy point defects, antisite pairs, clusters of defects, dislocations and disclinations, stacking faults, magnetic point, line and wall defects, $\cdots$)\\
A distinction is also made according to whether a process occurs without energy transfer ($\hbar \omega = 0$), in which case the scattering is said elastic, or takes place with an energy transfer in a window of finite width around zero, in which case the scattering is said quasi-elastic, or else gives rise to a finite energy transfer ($\hbar \omega \neq 0$), in which case the scattering is said inelastic. Coherent elastic neutron scattering emerges from the static correlations $\lim_{t \to \infty} <\alpha_{n\nu}\alpha_{m\mu}(t)>$ between distinct scattering centers in the crystal. It is basically answerable to interference between scattered waves and produces a signal if and only if strict geometrical conditions are satisfied. 

Now, we remind that one experimentally understands by diffraction any scattering method by which one collects outgoing neutrons in the spatial directions for which one meets the geometrical conditions for coherent elastic scattering, but without energy discrimination. It is then characterized by a cross section integrated over the outgoing neutron kinetic energy: $\mathrm{d}\sigma / \mathrm{d}\Omega = \int (\partial^2\sigma / \partial \Omega \partial E_f) dE_f$. If the ingoing neutron kinetic energy is fixed then this amounts to integrate over the energy transfer $\hbar \omega$. The method provides a snapshot of the states and correlations in the crystal. Along the diffracting directions the elastic scattering due to constructive interference most often dominates and one gets a Bragg peak superimposed over a background associated with all the other scattering processes. The elastic signal is deduced by subtracting the measured background in nearby spatial directions where no elastic scattering contributes.

\subsubsection{Nuclear and magnetic structure factors}

Consider first only the nuclear scattering from a crystal with several Bravais lattices labelled by the symbol $\nu$ whose atoms are spatially located by the vectors $\vec R_{\nu n} = \vec r_{\nu} + \vec R_n$ where $\vec r_{\nu}$ defines a position in the unit cell and $\vec R_n$ a cell position. If the nuclear spins are unfrozen then the energy-integrated elastic scattering cross section, one often more simply calls the differential scattering cross-section, is computed for unpolarized neutrons as \cite{MarshallLovesey}
$${d\sigma_N\over d\Omega} = {(2\pi)^3\over V_c}N_c\sum_{\vec H}|F_N(\vec Q)|^2\delta(\vec Q-\vec H)$$
$N_c$ is the number of unit cells and $V_c$ the unit cell volume. $\vec H$ is a reciprocal lattice vector. $\vec H=h\vec a^*+k\vec b^*+l\vec c^*$, where $\vec a^*={2\pi\over V_c}\vec b\times\vec c$, $\vec b^*={2\pi\over V_c}\vec c\times\vec a$, and $\vec c^*={2\pi\over V_c}\vec a\times\vec b$ and $(h, k, l)$ is a triplet of integer coefficients called the Miller indices. The quantity
$$F_N(\vec Q)=\sum_{\nu}b_{\nu}~e^{-W_\nu(\vec Q)}~e^{i\vec Q.\vec r_{\nu}}$$
defines the nuclear structure factor. $b_\nu = \sum_\xi c_\xi \{(\Upsilon_\xi+1) b_\xi^{(+)} + \Upsilon_\xi b_\xi^{(-)}\} / (2\Upsilon_\xi+1)$ is the Fermi length, averaged over the random isotope distributions and random nuclear spin orientation ($c_\xi$ is the proportion of the $\xi-$isotope with nuclear spin $\Upsilon_\xi$ and $b^{(\pm)}$ the Fermi lengths associated with the neutron-nucleon total spin states $\Upsilon_\xi \pm 1/2$. $e^{-W_\nu(\vec Q)} = <e^{-\vec Q \cdot \{\vec r_\nu(t)-\vec r_\nu\}}>$ is the Debye-Waller factor associated with the atom $\nu$.

Consider next  only the magnetic scattering from a crystal in an ordered magnetic phase. The frozen arrangement of the magnetic moments display then spatial periodicity, which may coincide with the nuclear one (ferromagnets or in-cell antiferromagnets), be a rational multiple of it (commensurate structure) or not (incommensurate structure), comprise harmonics (structure squaring), combine symmetry related periodicities (multi$-\vec k$ structures), $\cdots$. Whatever the case one is able to characterize the spatial periodicity with the help of propagation vectors $\vec k$ (inside or at the surface of the first Brillouin zone). The magnetic moment distribution can be expanded over its Fourier components $\vec m_{\nu,\vec k}$ as
$$\vec \mu_{\nu n}=\sum_{\vec k} \vec m_{\nu,\vec k}e^{-i\vec k.\vec R_n}$$
The differential scattering cross section for unpolarized neutrons is computed as \cite{MarshallLovesey}
$${d\sigma_M\over d\Omega} = {(2\pi)^3\over V_c}N_c\sum_{\vec H}\sum_{\vec k}|\vec F_{M\perp}(\vec Q)|^2\delta(\vec Q-\vec H-\vec k)$$
where $\vec F_{M\perp}(\vec Q) = (\vec Q \times \vec F_M(\vec Q) \times \vec Q) / \vec Q^2$ and the complex vector quantity 
$$\vec F_M(\vec Q=\vec H+\vec k)=p\sum_{\nu}f_{\nu}(\vec Q)~\vec m_{\nu,\vec k}~e^{-W_\nu(\vec Q)}~e^{i\vec Q.\vec r_{\nu}}$$
defines the magnetic structure factor $\vec F_M(\vec Q)$. Note that in the general case, and since the magnetic moment is a real quantity, every $\vec k$ should be associated with $-\vec k$ with $\vec m_{\nu,-\vec k}=(\vec m_{\nu,\vec k})^*$. We shall ignore from now on the Debye-Waller factors $e^{-W_\nu(\vec Q)}$.

In the case of an helical magnetic structure with {\bf positive chirality}, whose magnetic moments are lying in a plane ($\vec u$, $\vec v$), where $\vec u$ and $\vec v$ are orthogonal unit vectors, while rotating anticlockwise about the $\vec k$ direction, the magnetic moments can be written as
\footnote{When $\mu_{1\nu}=\mu_{2\nu}$ the helical structure is circular, otherwise it is elliptic. When $\mu_{2\nu}$=0, the magnetic arrangement is sine-wave modulated (collinear magnetic moments).}:
$$\vec \mu_{\nu,n}=\mu_{1\nu}.\vec u\cos(\vec k.\vec R_n+\Phi_{\nu})+\mu_{2\nu}.\vec v\sin(\vec k.\vec R_n+\Phi_{\nu})$$
This distribution is described by two Fourier components associated to $\vec k$ and $-\vec k$:
$$\vec m^+_{\nu,\vec k}=[{\mu_{1\nu}.\vec u+i\mu_{2\nu}.\vec v\over2}]e^{-i\Phi_{\nu}}$$
$$\vec m^+_{\nu,-\vec k}=[{\mu_{1\nu}.\vec u-i\mu_{2\nu}.\vec v\over2}]e^{i\Phi_{\nu}}$$
with $\vec m^+_{\nu,\vec k}=(\vec m^+_{\nu,-\vec k})^*$. The magnetic structure factor writes
$$\vec F^+_M(\vec Q=\vec H\pm\vec k)=p\sum_{\nu}f_{\nu}(\vec Q)\vec m^+_{\nu,\pm\vec k}e^{i\vec Q.\vec r_{\nu}}=p\sum_{\nu}f_{\nu}(\vec Q)[{\mu_{1\nu}.\vec u\pm i\mu_{2\nu}.\vec v\over2}]e^{\mp i\Phi_{\nu}}e^{i\vec Q.\vec r_{\nu}}$$

In the case of an helical magnetic structure with a {\bf negative chirality} (clockwise sense of rotation of the magnetic moments), the magnetic moments and magnetic structure factor write
$$\vec \mu_{\nu,n}=\mu_{1\nu}.\vec u\cos(\vec k.\vec R_n+\Phi_{\nu})-\mu_{2\nu}.\vec v\sin(\vec k.\vec R_n+\Phi_{\nu})$$ 
$$\vec F^-_M(\vec Q=\vec H\pm\vec k)= p\sum_{\nu}f_{\nu}(\vec Q)\vec m^-_{\nu,\pm\vec k}e^{i\vec Q.\vec r_{\nu}}=p\sum_{\nu}f_{\nu}(\vec Q)[{\mu_{1\nu}.\vec u\mp i\mu_{2\nu}.\vec v\over2}]e^{\mp i\Phi_{\nu}}e^{i\vec Q.\vec r_{\nu}}$$
$\vec k$ and -$\vec k$ correspond to positive and negative chirality respectively in the expression of the magnetic moments for a helical structure. Both structures have $\pm \vec k$ Fourier components and yield Bragg reflections at $\vec H\pm\vec k$. If we have one Bravais lattice ($\nu=1$) then $\vec m^+_{\nu,\vec k}=\vec m^-_{\nu,-\vec k}$. This is no more the case with several Bravais lattices.

The formalism is similar for a helix or for a cycloid. Triangular arrangements of coplanar spins at 120$^{\circ}$ of each other can be described as an helix. This is for instance the case of the antiferrochiral magnetic arrangement in the trigonal  compound RbFe(MoO$_4$)$_2$ with (1/3~1/3) propagation vector in the plane of the spins \cite{Kenzelmann2007}. The 120$^{\circ}$ spin arrangement can alternatively be obtained from a suited dephasing of the three Bravais lattices. An example is provided by the ferrochiral arrangement of the magnetic moments on the triangular network of iron triangles in the Fe langasite \cite{Marty2008} (see section \ref{langasite}). The outcomes for the chiral diffraction are then completely different.

\subsubsection{Chiral diffraction from a magnetic helix}

\begin{figure}
\resizebox{0.75\columnwidth}{!}{
\includegraphics{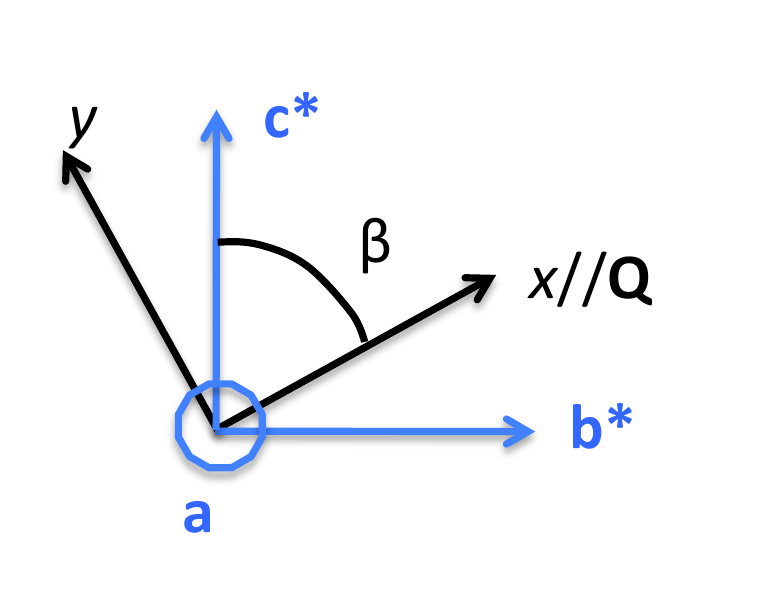} }
\caption{The scattering plane ($\vec b^*$, $\vec c^*$) for the considered helical arrangement with the spins lying in the ($\vec a$, $\vec b$) plane. The chiral vector $\vec \Xi$ is parallel to $\vec c^*$ and makes an angle $\beta$ with the scattering vector. The $x$ and $y$ axes of the right-handed frame used in neutron polarimetry are also shown, the $z$ axis being along $\vec a$.}
\label{fig:3}       
\end{figure}

Let us consider a magnetic arrangement, for instance on a trigonal lattice, with several Bravais lattices: the magnetic moments lie in the ($\vec a$, $\vec b$) plane and propagate as a circular helix along the $\vec c$ axis with a propagation vector $\vec k$. The scattering plane is ($\vec b^*$, $\vec c^*$), the vertical axis is $\vec a$, and $\beta$ is the angle between the scattering vector $\vec Q$ and the chiral vector $\vec \Xi=\vec u\times\vec v$ (see figure \ref{fig:3}).

The magnetic structure factor $\vec F_M(\vec Q)=(M_{\vec a},M_{\vec b^*},M_{\vec c^*})$ in the ($\vec a$, $\vec b^*$,$\vec c^*$) frame can be expressed in the conventional ($X$, $Y$, $Z$) frame as $(M_{\vec c^*}\cos\beta+M_{\vec b^*}\sin\beta,M_{\vec c^*}\sin\beta-M_{\vec b^*}\cos\beta,M_{\vec a})$. This leads to:
\vskip 0.2 cm
$\sigma^y_M=M_{\vec c^*}^{\dag}M_{\vec c^*}\sin^2\beta+M_{\vec b^*}^{\dag}M_{\vec b^*}\cos^2\beta-2\Re(M_{\vec c^*}^{\dag}M_{\vec b^*})\sin\beta\cos\beta$
\vskip 0.2 cm
$\sigma^z_M=M_{\vec a}^{\dag}M_{\vec a}$
\vskip 0.2 cm
$M_{ch}=2\sin\beta\Im(M_{\vec a}^{\dag}M_{\vec c^*})-2\cos\beta\Im(M_{\vec a}^{\dag}M_{\vec b^*})$
\vskip 0.2 cm
\noindent Now, setting $\vec u$ parallel to $\vec a$ and $\vec v$ parallel to $\vec b^*$ we write $F_M(\vec H\pm \vec k)={\vec u\pm \epsilon i\vec v\over 2}T$, 
where the scalar 
$$T=\sum_{\nu}f_{\nu}(\vec H\pm \vec k)\mu_{\nu}e^{\mp i\epsilon\phi_{\nu}}e^{i(\vec H\pm\vec k).\vec r_{\nu}}$$ 
depends on the atomic positions and dephasings of the different Bravais lattices and $\epsilon=\pm 1$ determines the chirality of the helix.
\vskip 0.2 cm
\noindent With these notations,
\vskip 0.2 cm
$\sigma^y_M=T^{\dag}T\cos^2\beta/4$
\vskip 0.2 cm
$\sigma^z_M=T^{\dag}T/4$ and 
\vskip 0.2 cm
$M_{ch}=\mp 2\epsilon\cos\beta T^{\dag}T/4$
\vskip 0.2 cm
\noindent from which one deduces the polarization matrix elements
$$P_{yx}=P_{zx}=\pm{2\epsilon\cos\beta\over \cos^2\beta+1}$$
\vskip 0.2 cm
\noindent the sign of which depends on the satellites $\pm \vec k$, on the angle $\beta$, and on the chirality sign $\epsilon$. One finds the opposite signs for $M_{ch}$ and for $P_{yx}=P_{zx}$ for opposite spin chirality.
\vskip 0.2 cm
\noindent
The differential cross-section writes
$${d\sigma_M\over d\Omega}(\vec Q=\vec H\pm \vec k)={(2\pi)^3\over V}N{T^{\dag}T\over 4}[1+\cos^2\beta\mp2\epsilon\cos\beta(\vec P_0.{\vec{\hat Q}})]\delta(\vec Q-\vec H\mp\vec k)$$
\vskip 0.2 cm
\noindent with $\vec{\hat Q}=\vec Q/|Q|$. For small $\beta$ value and $\vec Q$ (anti)parallel to the polarization $P_0$, a homochiral helix will indeed scatter only one polarization state of the neutrons, indicating the chirality of the helix. Such an experiment was performed on MnSi by Ishida {\it et al.}. It allowed deducing the left-handed chirality of the magnetic helix in the probed crystal \cite{Ishida1985}. Note that this scattering would not change if the crystal were rotated by 180$^{\circ}$ around the vertical axis indicating that the Friedel law remains valid for both satellites as a signature of P-odd processes. For centrosymmetric compounds with equally populated spin chirality domains, the chiral scattering term will average to zero. To observe some chiral scattering, an unbalanced spin chirality domain distribution must be achieved. The difference of opposite spin chirality volumes can be directly probed, replacing $\epsilon$ by the normalized volumes of right minus left chiralities in the above expressions. The above calculations of $M_{ch}$, $P_{yx}$, $P_{yz}$ and ${d\sigma_M\over d\Omega}$ with polarized neutrons also demonstrate that the chiral scattering term can probe the absolute spin chirality of a helix described by a propagation vector but is not sensitive to the absolute spin chirality associated with the spin dephasing between several Bravais lattices.

\subsection{Dynamical chirality}

In the previous section, we have computed the chiral scattering produced by magnetic orders presenting an unbalanced domain distribution of static spin chirality. This is related to the broken P-symmetry, intrinsic to non-centrosymmetric crystals or induced by external constraints in centrosymmetric ones with a non-centrosymmetric magnetic space group. These constraints can be electric/magnetic fields, elastic torsion, or defect pinning of the domains walls \cite{Gukasov1999}.

\begin{figure}
\resizebox{1.0\columnwidth}{!}{
\includegraphics{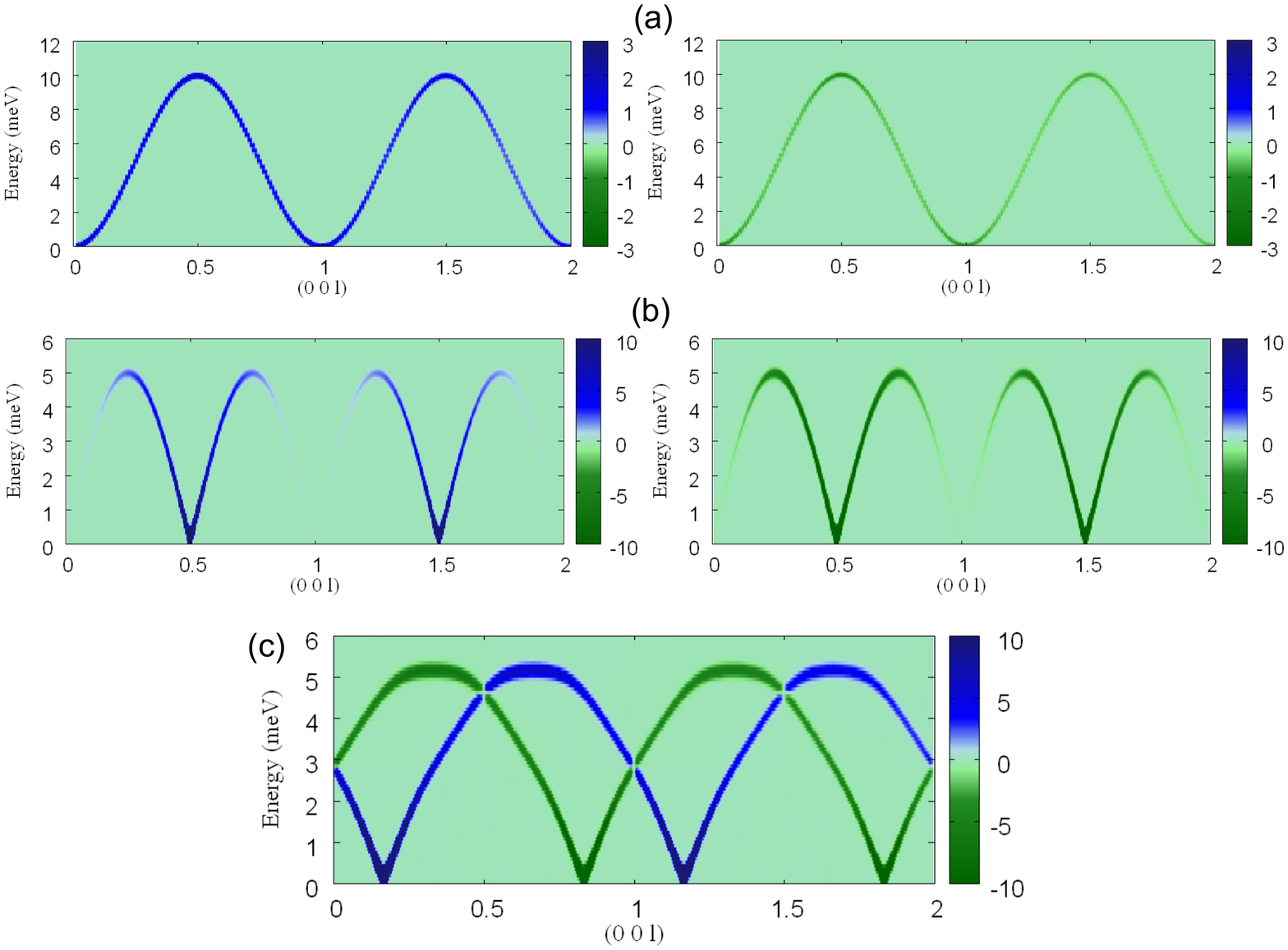} }
\caption{The calculated chiral scattering using the Holstein-Primakov formalism in the linear approximation of the spin waves emerging from (a) each single-domain of a ferromagnet, (b) of a collinear antiferromagnet with alternating spins along $z$ and from (c) a magnetic helix whose spins are rotating about the $z$ axis ($k$=1/6, $J_1/J_2$=-2) with a single chirality and planar anisotropy.}
\label{fig:4}       
\end{figure}

Another source of chiral scattering is the one associated with the intrinsic chirality of the spin wave excitations emerging from the ground state of an ordered phase (spin precession around the quantization axis associated with locally frozen moment). It is observed in the dynamical regime ($\omega\neq 0$) and exists even in magnetic materials that do not break the P-symmetry. The simplest example is that of a ferromagnet under an applied magnetic field, which thus breaks the T-symmetry macroscopically (single domain). The coherent inelastic cross-section is computed as \cite{MarshallLovesey}
$${d^2\sigma\over d\Omega dE_f}=$$
$${k_f\over k_i}{(2\pi)^3\over V}f^2(\vec Q)\mu^2\{[1+(\vec{\hat Q}.\vec{\hat M})^2-(\vec{\hat Q}.\vec{\hat M})(\vec{\hat Q}.\vec P_0)<n_q+1>]\delta(\hbar\omega-\epsilon_q)\delta(\vec Q-\vec H+\vec q)$$
$$+[1+(\vec{\vec{\hat Q}}.\vec{\hat M})^2+(\vec{\hat Q}.\vec{\hat M})(\vec{\hat Q}.\vec P_0)<n_q>]\delta(\hbar\omega+\epsilon_q)\delta(\vec Q-\vec H-\vec q)\}$$
where $\vec{\hat M}=\vec M/|M|$ with $\vec M$ the magnetization, $\vec q$ is the spin wave linear momentum and $\epsilon_q$ is the spin wave energy. $<n_q> = \{e^{\hbar \epsilon_{\vec q}}-1\}^{-1}$ is the Bose factor. The pure inelastic part of the scattering cross-section for a ferromagnet is polarization dependent. As in the case of a helical structure, with $\vec{\hat Q}//\vec P_0$, the magnon creation (absorption) is  possible only for one neutron polarization state of the incoming beam, the one for which the vectors $\vec M$ and $P_0$ are antiparallel (parallel). It should be emphasized that this chiral scattering is purely dynamical and usually disappears at $\omega=0$. It can be observed in compounds that do not break the space inversion (centrosymmetric) and stabilizes an achiral magnetic ground state. On the other hand, it requires T-symmetry breaking which is materialized by applying a magnetic field. The T-asymmetry can give rise to a finite chiral scattering integrated over $\omega$ when the magnetic field and the neutron inelastic scattering vector have a non-zero component along the wave-vector of the incoming neutrons. An enhancement of this asymmetry arises in small angle scattering \cite{Gukasov1999}. This dynamical chiral scattering of the spin waves is illustrated in figure \ref{fig:4}(a) where $M_{ch}$ has been computed using the standard Holstein-Primakov formalism in the linear approximation \cite{Holstein1940} for the two T-symmetry related ferromagnetic domains (software of S. Petit, Laboratoire L\'eon Brillouin). A fully chiral spin waves spectrum is observed with opposite chirality for the two domains. In the case of a collinear antiferromagnet, the dynamical chiral scattering disappears at the zone center $q=0$ but is observed at the zone boundary $q=\pi/2$ with an opposite sign for the two T-symmetry related domains (see figure \ref{fig:4}(b)), similarly to the ferromagnetic case. 

When the magnetic ground state is itself chiral the mechanisms by which the excitations give rise to a chiral scattering are less immediate to discern owing to the magnetic structure complexity. The case of the 120$^{\circ}$ spin configuration with staggered chirality in the quantum triangular antiferromagnet CsCuCl$_3$ has been studied by Syromyatnikov \cite{Syromyatnikov2005}. The magnetic structure of this centrosymmetric compound is described by a propagation vector (0 ~2/3 ~0) and exhibits a finite dynamical chiral scattering near the satellites if the chirality domains of the ground state are unbalanced, even in zero magnetic field. As another example, the dynamical chiral scattering of a helix propagating along the $z$-axis with one Bravais lattice and a single-domain chirality, was calculated within the Holstein-Primakov formalism in the linear approximation (see figure \ref{fig:4}(c)). $M_{ch}$ is finite when emerging from the magnetic satellites and up to the maximum energy with opposite sign for the $+-$  magnetic satellites. $M_{ch}$ cancels at the zone center. This dynamical chiral scattering due to spin waves emerging from a chiral magnetic ground state has been observed experimentally and confirmed by calculations in the Fe-langasite, as described in the next section \cite{Loire2011,Jensen2011}.

A topic of utmost interest was the possibility to probe chirality fluctuations, in particular since Kawamura predicted a new universality class specific to chiral criticality \cite{Kawamura1998}. This in principle would require to measure correlation functions involving at least four spins, which cannot be accessed by neutron scattering. Maleyev however suggested that these multi-spin correlations could be detected though their projection on any axial vector that would be present in the system of spins (magnetic field, macroscopic magnetization, DM vector interaction or the spin chirality vector characterizing the ground state) \cite{Maleyev1995,Syromyatnikov2005}. The proposed experimental protocol was implemented for several chiral compounds \cite{Plathky1998}, though it was to some extent questioned \cite{Lorenzo2007}. It was in particular argued that no clear-cut distinction might be made between a chirality, that even magnons in a ferromagnet would show, and a more exotic chirality that for instance would characterize excitations and correlations in systems showing multi-spin orders with zero on-site spin average \cite{Chandra1991,Gorkov1992}. One must here face again the subtleties inherent to the dynamical aspects of the chirality.

\section{Case study: the Fe Langasite}
\label{langasite}

\begin{figure}
\resizebox{1.1\columnwidth}{!}{
\includegraphics{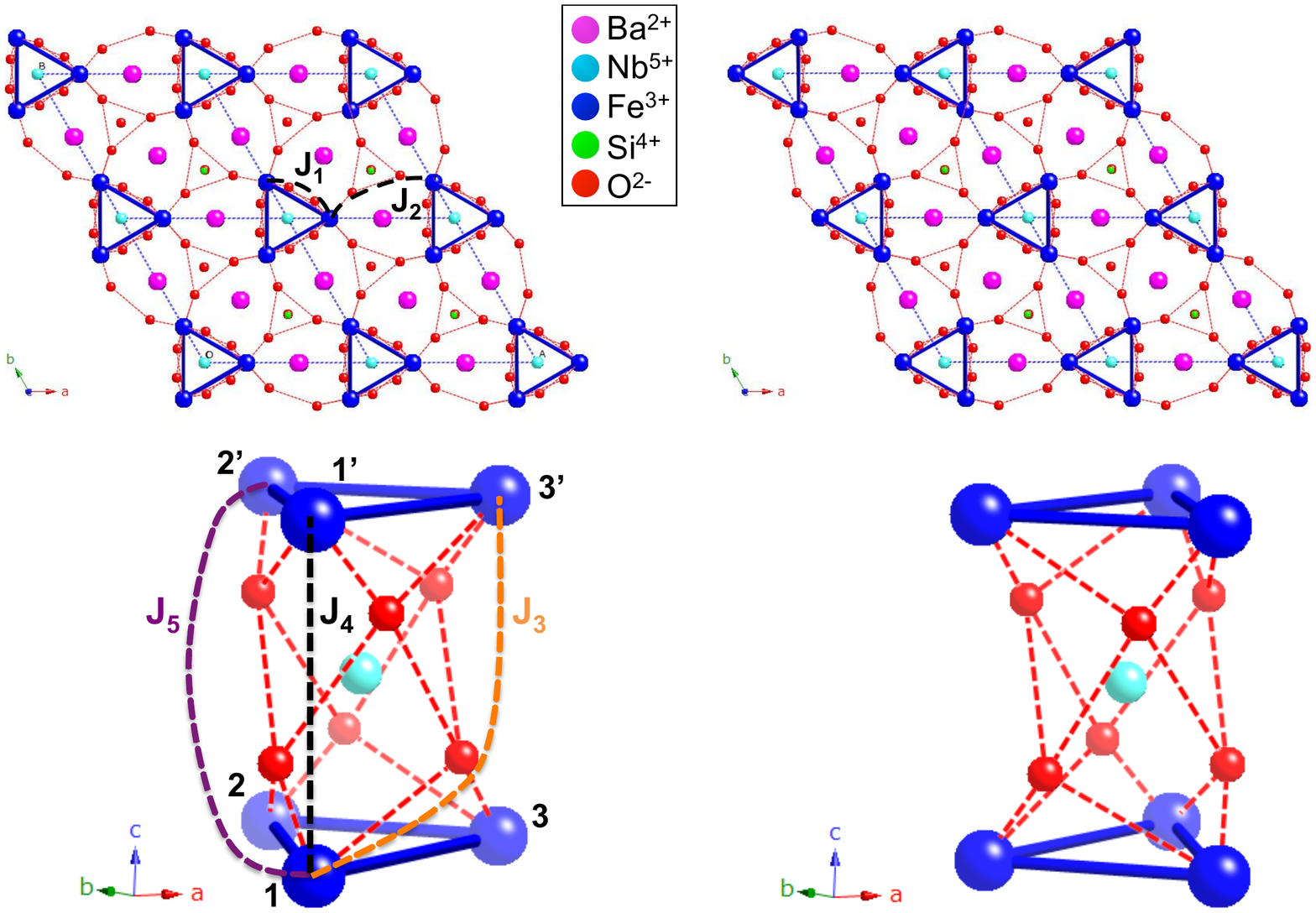} }
\caption{The structure, projected in the ($\vec a$, $\vec b$) plane and viewed as stacked along the $\vec c$ axis, of the two Ba$_3$NbFe$_3$Si$_2$O$_{14}$ enantiomorphs with negative (left) and positive (right) structural chiralities. The 5 exchange paths $J_1$ to $J_5$ mediated by oxygens are shown as dashed lines.}
\label{fig:5}       
\end{figure}

In this section, we consider more specifically the chiral properties, nuclear/magnetic, static/dynamical, of an iron oxide belonging to the langasite family, Ba$_3$NbFe$_3$Si$_2$O$_{14}$. This material crystallizes in the non-centrosymmetric trigonal space group $P$321 with lattice parameter $a$=8.539 and $c$=5.524 \AA. The structure is chiral and Ba$_3$NbFe$_3$Si$_2$O$_{14}$ can be synthesized in two enantiomorphic species. The magnetic Fe$^{3+}$ ions (with spin $S$=5/2 and orbital angular momentum $L$=0) at the Wyckoff position $3f$ form a triangular array of triangles (trimer units) in the ($a$, $b$) plane (see figure \ref{fig:5}). From magnetization and specific heat measurements, a transition to an antiferromagnetic ordered phase was evidenced at T$_N$=27 K \cite{Marty2008,Marty2009}. Magnetodielectric properties and a possible ferroelectric polarization appear concomitantly \cite{Marty2010,Zhou2009}. Other Fe langasite have been investigated with various chemical substitution on the non-magnetic sites and exhibit rather similar properties \cite{Marty2010,Yu2009}. The remarkable magnetic behavior of these compounds have attracted a lot of interest recently \cite{Stock2011,Zhou2010,Lee2010,Lyubutin2011,Zorko2011}. We report below our investigation of the magnetic ground state and the excitations of this material by neutron scattering, using polarized neutrons and polarization analysis \cite{Marty2008,Loire2011}.

\subsection{Static chirality}

\begin{figure}
\resizebox{1\columnwidth}{!}{
\includegraphics{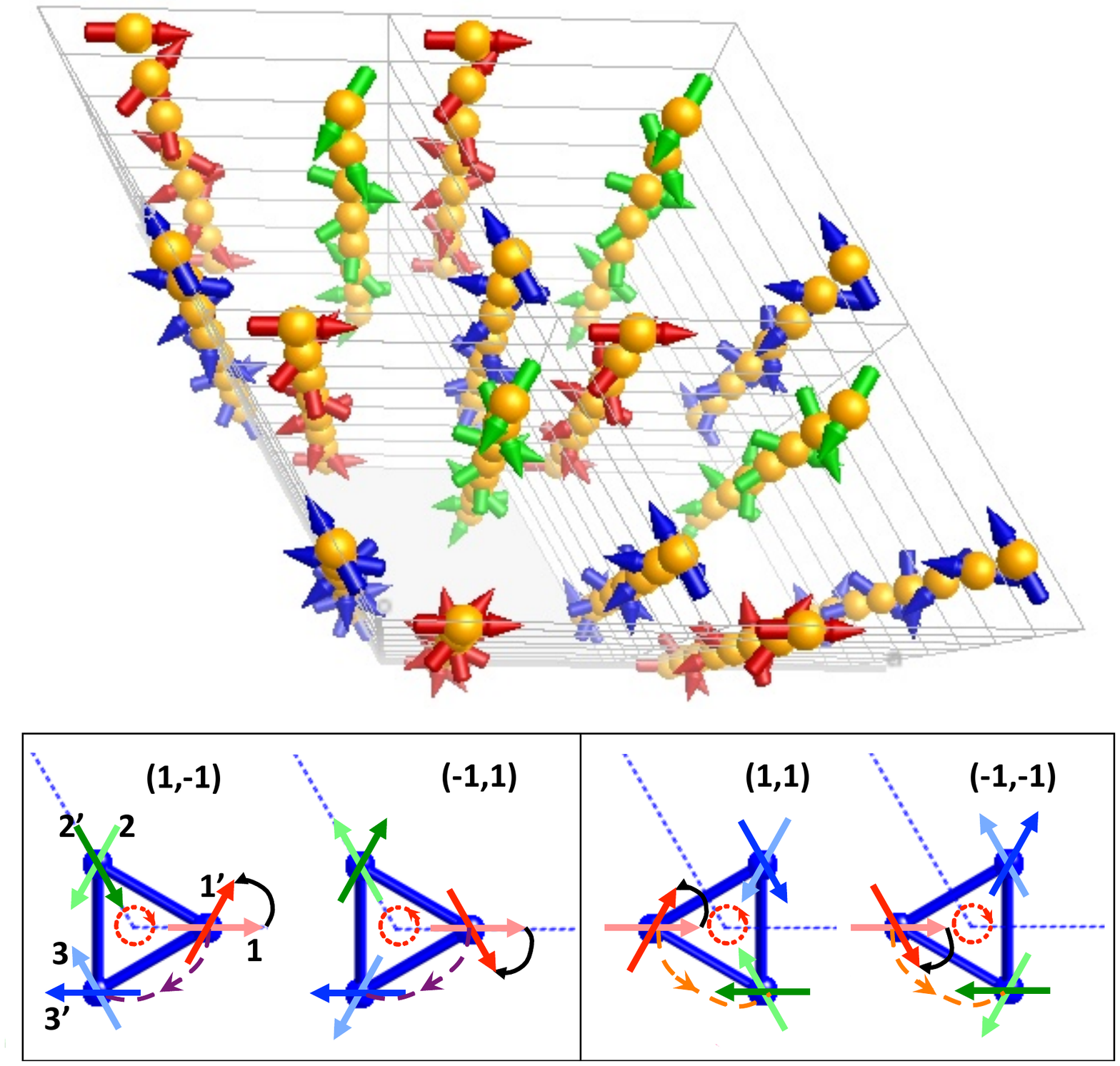} }
\caption{Top : Magnetic structure of Ba$_3$NbFe$_3$Si$_2$O$_{14}$ with different colors for the three Bravais lattices. Below : Representation of the magnetic structures associated with the 4 possible chiral ground states (helical chirality=$\pm 1$, triangular chirality=$\pm 1$). The light colored moments lie in one layer and the darker colored ones in the next layer along the $\vec c$ axis, a black curved arrow defines the helical chirality. The red arrowed circle materializes the triangular chirality. The structural chirality is related to the strongest diagonal exchange between the two layers, which is shown as a purple/orange dashed arrow path for negative/positive structural chirality.}
\label{fig:6}       
\end{figure}

As a first step, we performed a powder neutron diffraction experiment on the D1B (CRG-CNRS-ILL) diffractometer that showed the rise of extra Bragg peaks below T$_N$, which could be indexed with the propagation vector $\vec k=(0,0,\approx1/7)$. The symmetry analysis of the little group associated to this propagation vector provided three possible arrangements of the 3 Bravais lattices of spins, all associated to irreducible representations of order 1. The character table and basis vectors are reported in table \ref{representation}. The second representation corresponds to a ferromagnetic alignment in the ($\vec a$,$\vec b$) plane of the three basis vectors. The first and third representations yield a positive and negative triangular chirality respectively: the three magnetic moments of a trimer are orientated at 120$^{\circ}$ from each other in the ($\vec a$,$\vec b$) plane, rotating anticlockwise ($\tau_1$) and clockwise ($\tau_3$). A component along the $\vec c$-axis is allowed in all cases. Refinement of the magnetic intensities obtained from powder and single-crystal neutron diffraction (D1B and D15, ILL) agree with representations 1 or 3. The 120$^{\circ}$ spin arrangement in the ($\vec a$,$\vec b$) plane is helically modulated in the perpendicular direction with a helix period of $7$ within the measurements accuracy (see figure \ref{fig:6}). A small out-of-plane component could also be present as suggested by magnetization measurements \cite{Marty2008}. This magnetic structure is thus characterized by two kinds of magnetic chiralities, related to the sense of rotation of the spins around the triangle (triangular chirality) and related to the sense of rotation of the spins around the helix axis (helical chirality), in addition to the structural chirality. Note that the symmetry analysis allows both triangular chiralities, each one being possibly associated to a positive ($+\vec k$) or negative helical chirality ($-\vec k$).

\begin{table}
\caption{Character table (top) and basis vectors (bottom) of the three irreducible representations $\tau_1$ to $\tau_3$, with $\epsilon=e^{i2\pi/3}$ and $x$=0.2496}
\label{representation}       
\begin{center}
\begin{tabular}{|c|ccc|}
\hline\noalign{\smallskip}
  & Id & 3$^+$ & 3$^-$ \\
  & ($x,y,z$) & ($-y,x-y,z$) & ($-x+y,-x,z$) \\
\hline\noalign{\smallskip}
$\tau_1$ & 1 & 1 & 1 \\
$\tau_2$ & 1 & $\epsilon$ & $\epsilon^2$ \\
$\tau_3$ & 1 & $\epsilon^2$ & $\epsilon$ \\
\hline\noalign{\smallskip}
 & Fe$_1$ & Fe$_2$ & Fe$_3$ \\
 &($x$,0,0.5)&(0,$x$,0.5)&(-$x$,-$x$,0.5) \\
\hline\noalign{\smallskip}
$\tau_1$ &  $u,v,w$ & $-v,u-v,w$ & $-u+v,-u,w$ \\
$\tau_2$ &  $u,v,w$ & $\epsilon^2(-v,u-v,w)$ & $\epsilon(-u+v,-u,w)$\\
$\tau_3$ &  $u,v,w$ & $\epsilon(-v,u-v,w)$ & $\epsilon^2(-u+v,-u,w)$\\
\noalign{\smallskip}\hline
\end{tabular}
\end{center}
\end{table}

The powder neutron diffraction experiment ruled out the ferromagnetic solution but did not allow to distinguish between the different 120$^{\circ}$ magnetic arrangements and to get information about the magnetic chirality. Experiments on a single-crystal were necessary to achieve this. First, the structural chirality of the single-crystal used during the neutron experiments was determined using the anomalous x-ray diffraction \footnote{Note that, for this special case, the two enantiomorphic crystals  interconverted into each other by the inversion center can be equivalently conceived as being interchanged by the $z\rightarrow -z$ coordinate transform followed by a 60$^{\circ}-$rotation of the unit cell in the ($a$,$b$) plane.}. This has important consequences on the geometry of the magnetic exchange paths between the Fe$^{3+}$ moments mediated by oxygen atoms (see figure \ref{fig:5}). The nearest-neighbor antiferromagnetic interaction between the magnetic moments within the trimers, $J_1$, and the second-neighbor antiferromagnetic interaction within the plane connecting one moment of one trimers to 4 moments of the neighboring trimers, $J_2$, are equivalent for both structural chiralities. There are three additional antiferromagnetic interactions, $J_3$, $J_4$, and $J_5$, connecting the three atoms of one trimer to those of the superimposed trimers along the $\vec c$-axis. Due to the lack of inversion center, these 3 interactions are not equal (different geometry of the exchange paths) with, in particular, one diagonal interaction (either $J_3$ or $J_5$) expected to be stronger than the other two (\textit{i.e.} $J_3 >$ max($J_4$,$J_5$) or $J_5 >$ max($J_3$,$J_4$)). This leads to a twisted magnetic exchange around the superposed trimers. The sense of the torsion of these exchange paths is opposite in the two enantiomers since $J_3$ is changed to $J_5$ and {\it vice versa}. We will in the following arbitrarily call the structural chirality positive (negative) for a strong $J_3$ ($J_5$).

Given the structural chirality, neutron diffraction on a single crystal shows that the ground state is compatible with two magnetic chiralities out of four possibilities. These are given by the couple (helical chirality=$\pm 1$, triangular chirality=$\pm 1$) combining the two states $\pm1$ of the two magnetic chiralities (anticlockwise ($+1$) and clockwise ($-1$) senses of rotation of the magnetic moments). The integrated intensities obtained on a crystal with a positive (resp. negative) structural chirality agree only with the (+1,+1) or (-1,-1) (resp. (+1,-1) or (-1,+1)) chiral ground states. A simple picture, considering the 120$^{\circ}$ spin arrangement on a trimer ($J_1$) and the strongest out-of-plane interaction ($J_5$ or $J_3$) that connect it to the superposed trimer, allows to understand the relation between the structural and the magnetic chiralities. Due to the out-of-plane interaction, one magnetic moment of a given trimer will be anti-aligned with the magnetic moment of the upper trimer in the diagonal direction (e.g., atoms 1 and 3'  in figure \ref{fig:6}). This will result in a 60$^{\circ}$ rotation of the spins around the $\vec c$ axis (e.g., from atom 1 to atom 1' in figure \ref{fig:6}) leading to a propagation vector (0, 0, 1/6), and in the conservation of the triangular chirality on the upper trimer. If the diagonal interaction is twisted in the other sense, i.e. for the opposite structural chirality, the sense of rotation of the spins around the $\vec c$ axis will be opposite. The helical chirality is thus related to the triangular chirality via the clockwise/anticlockwise torsion of the exchange paths along the $\vec c$ axis resulting from the structural chirality. The finite value of the two weaker competing out-of-plane interactions further modulates the periodicity of the helix (additional $J_4=7\%~J_5$ and $J_3=22\%~J_5$ yield for instance a (0, 0, 1/7) propagation vector). This mechanism was confirmed by mean field calculation at zero temperature \cite{Bertaut} taking into account the 5 exchange paths. Given the structural chirality (choice of strong $J_3$ or $J_5$), the diagonalization of the Fourier transform of the interaction matrix yields three solutions characterized by the propagation vector (0, 0, $k$): a less favored ferromagnetic arrangement of the three Bravais lattices and the two degenerate 120$^{\circ}$ solutions of opposite triangular chirality. Figure \ref{fig:6} summarizes the interplay of the structural and magnetic chiralities leading to the four possible magnetic ground states.  

Finally, we made use of spherical polarimetry (CRYOPAD on the triple-axis IN22-CRG-CEA spectrometer at ILL) to further check the magnetic chiral state of the Fe langasite in the ordered phase. We collected the spin flip and non-spin flip intensities in three orthogonal directions independently for the incoming and scattered beam on 8 magnetic peaks of the type (-1,2,$l\pm k$) and (1,-2,$l\pm k$) with $l\in[0, 3]$. We showed in section \ref{chiraldiff} that the polarization matrices are only sensitive to the helical chirality, with the components $P_{yx}$ and $P_{zx}$ being proportional to the associated distribution of chirality domains \footnote{The calculations presented for a scattering plane ($\vec b^*$, $\vec c^*$) yield identical results for the ($\vec b$, $\vec c^*$) scattering plane used in the experiment.}. The sign of $P_{yx}$ and $P_{zx}$ indicated a positive helicity for our right-handed crystal and a fit of our data with respect to domain proportions systematically led to a single helical chirality domain in the crystal (see Table \ref{Pyx}). This, in turn, proved the selection of one (+1,+1) chirality couple, to agree with the unpolarized diffraction data, thus the selection of the unique associated positive triangular chirality. The same experiment was performed on a Ba$_3$NbFe$_3$Si$_2$O$_{14}$ crystal with a negative structural chirality. This led to the opposite helical chirality, being thus compatible with the (-1,+1) chiral ground state. 

\begin{table}
\caption{$P_{yx}$ and $P_{zx}$ polarization matrix elements for 8 magnetic Bragg peaks measured at 5 K on a Ba$_3$NbFe$_3$Si$_2$O$_{14}$ crystal with positive structural chirality by spherical polarimetry on IN22 with $\lambda$= 2.36 \AA. The scattering plane is ($\vec b$,$\vec c^*$).}
\label{Pyx}       
\begin{tabular}{c|c|c}
\hline\noalign{\smallskip}
Bragg &$P_{yx}$&$P_{zx}$\\
\hline\noalign{\smallskip}
 (-1 2 1+$\tau$) & -0.8185 (47) & -0.811 (5) \\
 (-1 2 -1-$\tau$) & -0.776 (5) & -0.799 (5) \\
 (-1 2 0+$\tau$) & -0.173 (6) & -0.213 (6) \\
 (-1 2 0-$\tau$) & -0.175 (6) & -0.218 (6) \\
 (-1 2 2+$\tau$) & -0.891 (17) & -0.859 (17) \\
 (-1 2 3+$\tau$) & -0.883 (18) & -0.845 (19) \\
 (-1 2 1-$\tau$) & 0.777 (2) & 0.7565 (23) \\
 (1 -2 2+$\tau$) & -0.863 (9) & -0.846 (9) \\
\noalign{\smallskip}\hline
\end{tabular}
\end{table}

To explain this ultimate selection, an additional mechanism must be invoked. We proposed the presence of the Dzyaloshinskii-Moriya (DM) antisymmetric exchange interaction $\vec D. (\vec S_i\times\vec S_j)$ with $\vec D$ the DM vector, always allowed in the absence of an inversion center between spins $\vec S_i$ and $\vec S_j$. It suffices to consider the DM interaction inside the trimer with the same DM vector along the $\vec c$ axis for the three bonds. This favors planar spin components, and its sign selects a triangular chirality and hence a helicity. This picture has been validated by electron spin resonance (ESR) measurements \cite{Zorko2011}. The ESR measurements also prove that another component of the DM interaction is present which corresponds to the DM vector perpendicular to the trimer bonds and lying in the ($\vec a$,$\vec b$) plane. It is expected to give rise to the small out-of-plane component suggested by the magnetization measurements. The conservation of the triangular chirality +1 for both structural chiralities is consistent with the P-even symmetry of the DM axial vector. 

\subsection{Dynamical chirality}

The remarkable ground state of the non-centrosymmetric Ba$_3$NbFe$_3$Si$_2$O$_{14}$ langasite exhibiting a pure chiral state combining two types of magnetic chiralities, urged us to study its dynamical properties. We were in particular interested in investigating the consequence of this chiral ground state on the magnetic excitations. We thus conducted experiments on the triple-axis spectrometer IN20 at the ILL in its polarized neutron setup, with polarizing Heusler crystals as monochromator and analyzer. We used the CRYOPAD device, to obtain strict zero-field environment at the sample position, and to prepare incoming and outgoing neutron polarization independently. We mainly used longitudinal polarization analysis in order to measure the dynamical chiral scattering cross-section $M_{ch}$ in addition to the dynamical magnetic cross-section $\sigma^y_M+\sigma^z_M$ (see section \ref{longitudinal}). The crystal, with a negative structural chirality, was aligned with ($\vec b^*$, $\vec c^*$) as a scattering plane, $\vec a$ being vertical. To get first a global overview of the spin waves, an experiment using unpolarized neutrons was also performed on the time-of-flight spectrometer IN5 at the ILL on the same single crystal at successive rotating angles around the vertical axis $\vec a$ \cite{Loire2011}. 

\begin{figure}
\resizebox{1\columnwidth}{!}{
\includegraphics{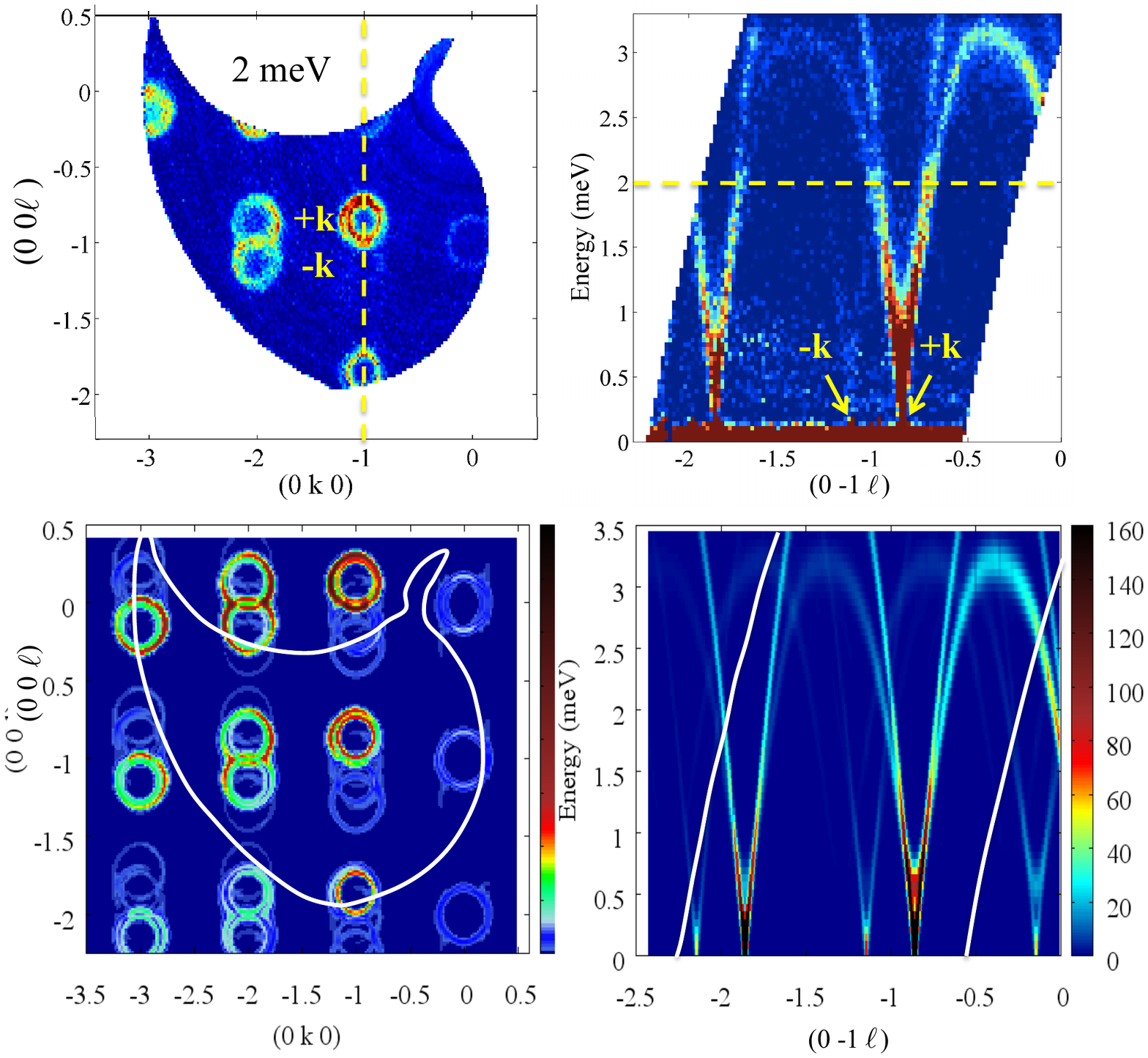} }
\caption{Spin waves measured at 1.5 K and 4 \AA\ with unpolarized neutrons on IN5 (top) and compared to calculations (bottom) at constant energy (left) and along the (0,-1,$\ell$) reciprocal space direction. The yellow dashed lines materialize the $\vec Q$-cut and Energy-cut shown in the alongside figure. The white lines show the reciprocal space area probed during the experiment.}
\label{fig:7}       
\end{figure}

\begin{figure}
\resizebox{1.0\columnwidth}{!}{
\includegraphics{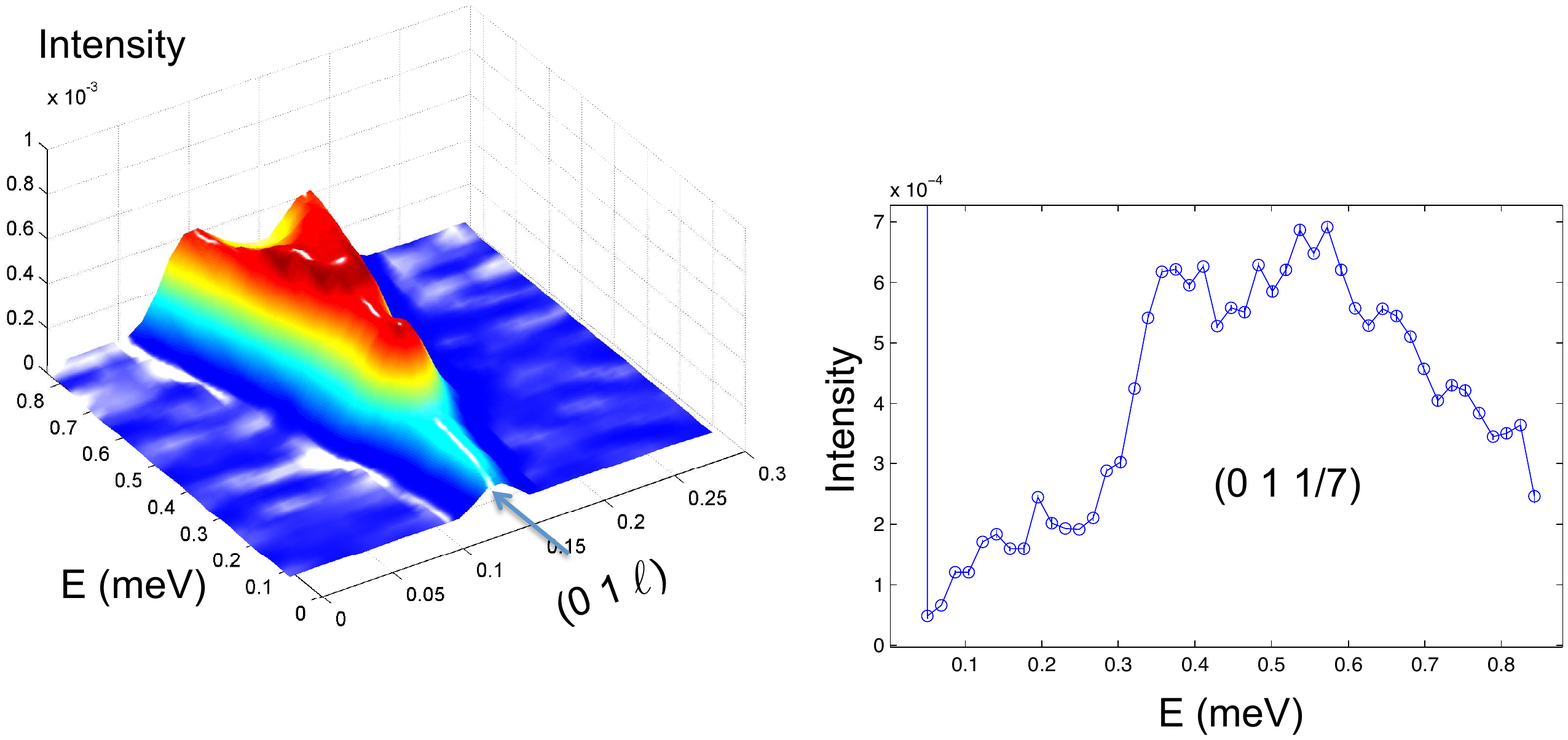} }
\caption{Spin gaps of $\approx$0.35 meV and possibly of 0.1 meV as measured on IN5 at 1.5 K at the magnetic satellite position with a wavelength of 8 \AA\ yielding an energy resolution of $\approx$0.025 meV. The right plot is a cut of the left figure at -0.025$<h<$0.025, 0.975$<k<$1.025, and 0.019$<\ell<$0.169. The error bars are within the dot size. The larger gap of the lower branch is attributed to the DM interaction whereas the origin of the smaller gap is not accounted for by the model hamiltonian described in the text and could originate from small single ion anisotropies.}
\label{fig:8}       
\end{figure}

From unpolarized neutron scattering, as shown in figure \ref{fig:7}, we observed two spin-wave branches emerging from the $\pm \vec k$ magnetic satellites that form delicate arches, with different maximum energies ($\approx$3.2 meV for the lower branch and $\approx$5 meV for the upper branch). One of the branches is gapped with a minimum at around 0.35~meV whereas the other branch might present a smaller gap of the order of 0.1~meV (see figure \ref{fig:8}). The intensities of the excitations emerging from the two magnetic satellites are very different around certain reciprocal lattice nodes, for instance along the line (0, -1, $\ell$). This is a consequence of the structural chirality as confirmed by spin waves calculations performed using the standard Holstein-Primakov formalism in the linear approximation with the software developed at the Laboratoire L\'eon Brillouin by S. Petit. A good agreement between experiment and calculation is achieved with the exchange parameters (in meV) $J_1=0.85\pm 0.1$, $J_2=0.24\pm 0.05$, $J_3=0.053\pm 0.03$, $J_4=0.017\pm 0.05$ and $J_5=0.24\pm 0.05$, constrained to fulfill the $k$=1/7 conditions. An additional DM vector along $\vec c$ of $\approx$1\%$|J_1|$ was checked to produce the lower branch gap of $\approx$0.35 meV and to select a triangular chirality. In the calculated dynamical magnetic cross-section, the asymmetric spectral weight of branches emerging from the +$\vec k$ and -$\vec k$ satellites is indeed inverted for opposite structural chiralities \footnote{Slightly different exchange parameters were obtained by J. Jensen in a calculation where the strongly coupled spin triangles are described as trimerized units placed in a mean field \cite{Jensen2011}.}. 

\begin{figure}
\resizebox{1.0\columnwidth}{!}{
\includegraphics{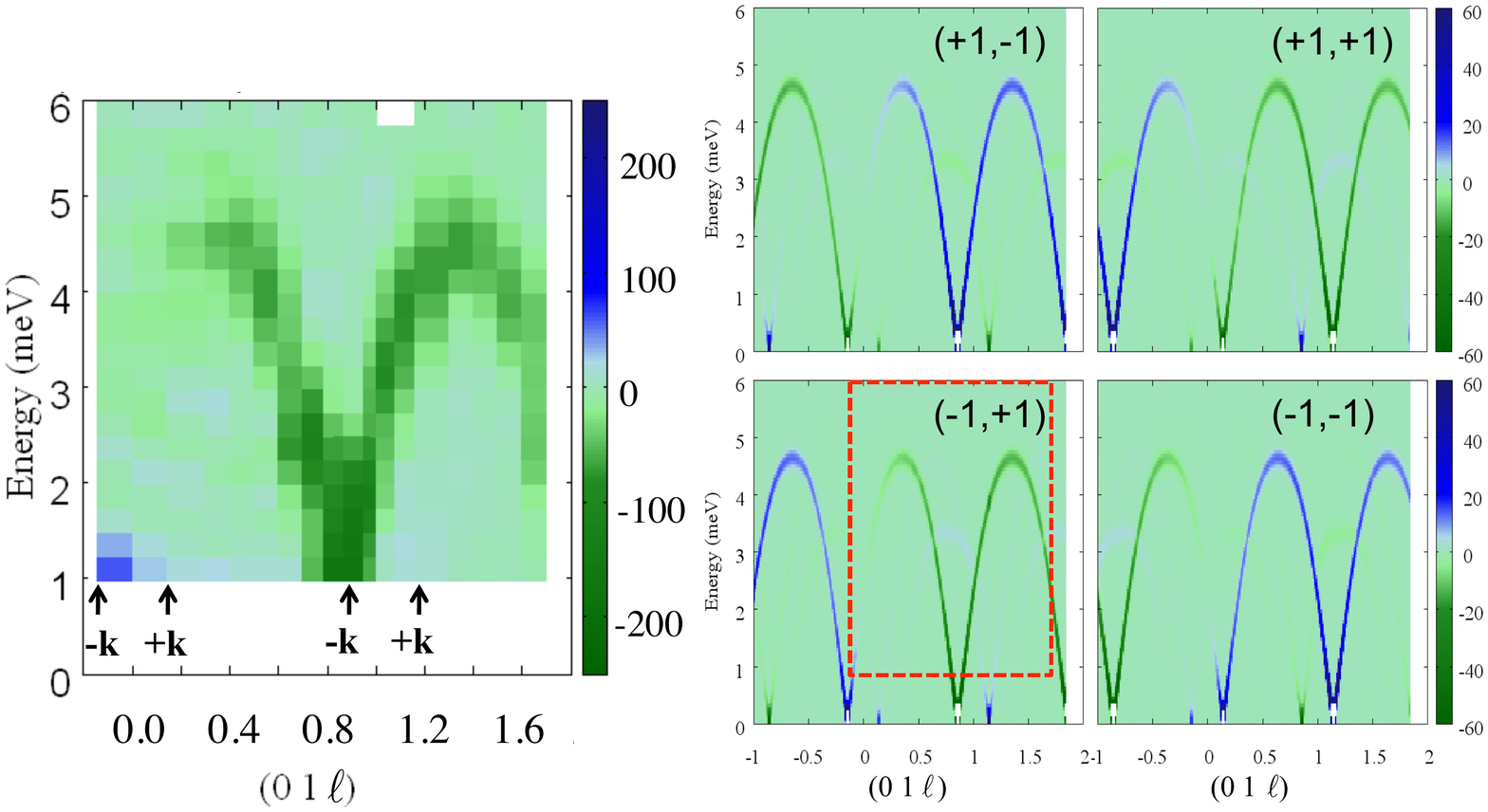} }
\caption{Chiral scattering of the spin waves measured (left) with polarized neutrons and polarization analysis on IN20 at 1.5 K with fixed $E_f$=14.7 meV. The results are compared to calculations (right) for the 4 possible chiral ground states ($\pm1$, $\pm1$). The red frame points out the solution agreeing with the experiment.}
\label{fig:9}       
\end{figure}
Even more interesting are the results obtained using neutron polarimetry. By extracting the chiral scattering contribution of the spin wave spectrum, we found that the lower mode is achiral whereas the upper mode has finite chirality (see figure \ref{fig:9}). As shown section \ref{chiraldiff}, the change of sign, from positive to negative, of the chiral scattering of the upper mode when $\ell$ becomes negative ($\beta$ greater than $\pi/2$ in figure \ref{fig:3}) is due to the fact that the neutron probes the moment component perpendicular to $\vec Q$ corresponding then to the projection of the Fourier transformed dynamical chirality onto $\vec Q$: $M_{ch}=\mp 2\epsilon\cos\beta T^{\dag}T/4$ for $\pm \vec k$ satellites with $\epsilon=\pm 1$ the helical chirality. This reflects a globally unchanged chirality of the upper spin wave branch over the whole energy spectrum. According to the calculations, this branch corresponds to correlation functions involving spin components along $\vec a$ and $\vec b$, and only along $\vec c$ for the lower branch. The latter yields a zero spin cross product and explains the absence of chiral scattering for the lower branch. The chiral scattering was also calculated and reproduces the measured one for the (-1,+1) chiral ground state \footnote{This is different from the previously published one \cite{Loire2011} due to a sign mistake in the experimental absolute chirality determination.}. Note that the spin wave chiral scattering is distinct for the 4 possible chiral ground states, either in the intensity or in the sign of the branches emerging from the two satellites. It thus provides us with a strong fingerprint of the ground state chirality. 

Our neutron polarimetry study of an enantiopure Fe langasite crystal thus revealed the fully chiral scattering of the spin waves emerging from a ground state presenting a single domain magnetic triangular and helical chiralities. This unique observation is a manifestation of the intrinsic chirality of the spin waves, enabled by the absence of chirality mixing in the ground state of this non-centrosymmetric compound \cite{Loire2011,Jensen2011}. The dynamical chiral scattering is in this case a strong fingerprint of the chiral ground state. Further work is ongoing concerning the chiral character of the paramagnetic fluctuations. 

\section{Conclusion}
In conclusion, although not exhaustive, neutron scattering is a very useful probe of the magnetic chirality. The dynamical chiral scattering probes the time and space Fourier Transforms of the dynamical antisymmetric spin correlation function, involving the cross product of the Fourier components of the magnetization perpendicular to the scattering vector. In the static case, this gives access to the projection onto the scattering vector of the chirality vector or spin current characterizing helical structures: helices or cycloids, and triangular spin arrangement described using a propagation vector. In the dynamical case, the intrinsic chirality of the spin waves can be accessed, whereas the experimental evidence of more exotic dynamical chirality, characteristic in particular of non ordered phases, is still an interesting challenge.

\acknowledgments About the authors and acknowledgments: Virginie Simonet and Rafik Ballou are working in the field of frustrated magnetism and multiferroism in particular through neutron scattering studies. During the two PhD works of K. Marty (post-doctoral position at SPINTEC, Grenoble, France) and M. Loire (now teacher in a secondary school in Nantes, France), they have been studying the chirality properties of a Fe langasite compound. This study has been impulsed by P. Bordet (Institut N\'eel) who stressed the interest of this compound in the context of frustrated magnetism and multiferroism. During the writing of his thesis manuscript, M. Loire noticed some inconsistencies between the magnetic chirality and the general definition of chirality which motivated this review article. P. Lejay, A. Hadj-Azzem, J. Balay, J. Debray, K. Marty and his supervisor P. Bordet have been deeply involved in the sample synthesis and preparation and in the study of the langasites at the Institut N\'eel. Various neutron experiments have been performed to unveil the fascinating static and dynamic properties of magnetic chirality of this material: powder neutron scattering on D1B (CRG-CNRS installed @ ILL) with O. Isnard (Institut N\'eel, Grenoble), single crystal diffraction on D15 and IN22 with CRYOPAD (CRG-CEA instruments installed @ ILL) led by E. Ressouche and with the help of L.-P. Regnault (CEA-Grenoble). Inelastic neutron scattering measurements have been obtained on ILL (Institut Laue Langevin, Grenoble, France) instruments with the help of ILL local contacts : IN5 with J. Ollivier and IN20 with M. Enderle and P. Steffens. The analysis of the spin waves was numerically performed using the software developed by S. Petit (Laboratoire L\'eon Brillouin, Gif-sur-Yvette, France) that computes  spin wave dispersions, spectral weights and various dynamical correlation functions within the linear approximation of the standard Holstein-Primakov formalism. The authors would like to thank E. Ressouche, S. Petit, L. C. Chapon (ILL), L.-P. Regnault and M. Enderle for very fruitful discussions about chirality. L. C. Chapon, S. Petit and E. Ressouche are also acknowledged for their careful reading of this manuscript and their very useful remarks. We wish to thank the ILL for the beam-time allocated. Financial support for this work was provided by the French Agence Nationale de la Recherche, Grant No. ANR-06- BLAN-01871.

\end{document}